\documentclass[pra,twocolumn,superscriptaddress,notitlepage,showpacs,showkeys]{revtex4-1}
\usepackage{graphicx,amsmath,amsfonts,amssymb,upgreek,txfonts,color,physics}
\usepackage[colorlinks,linkcolor=blue,citecolor=blue,urlcolor=blue,breaklinks=true]{hyperref}
\usepackage[utf8x]{inputenc}
\usepackage{color}
\usepackage{import}
\usepackage{enumitem}
\usepackage{lipsum}
\usepackage{mathtools}
\usepackage[dvipsnames]{xcolor}
\usepackage{subfigure}

\begin{document}

\title{Effects of quadratic optomechanical coupling on bipartite entanglements, mechanical ground-state cooling and squeezing in an electro-optomechanical system}

\author{N. Ghorbani}
\email{nafisph@gmail.com}
\address{Department of Physics, University of Isfahan, Hezar-Jerib, 81746-73441, Isfahan, Iran}

\author{Ali Motazedifard}
\email{motazedifard.ali@gmail.com}
\address{Quantum Sensing Lab, Quantum Metrology Group, Iranian Center for Quantum Technologies (ICQT), Tehran, Tehran 15998-14713, Iran}
\address{Quantum Optics Group, Department of Physics, University of Isfahan, Hezar-Jerib, Isfahan 81746-73441, Iran}

\author{M. H. Naderi}
\email{mhnaderi@sci.ui.ac.ir}
\address{Department of Physics, University of Isfahan, Hezar-Jerib, 81746-73441, Isfahan, Iran}
\address{Quantum Optics Group, Department of Physics, University of Isfahan, Hezar-Jerib, Isfahan 81746-73441, Iran}


\begin{abstract}
We theoretically investigate the steady-state bipartite entanglements, mechanical ground-state cooling, and mechanical quadrature squeezing in a hybrid electro-optomechanical system where a moving membrane is linearly coupled to the microwave field mode of an LC circuit, while it simultaneously interacts both linearly and quadratically with the radiation pressure of a single-mode optical cavity. We show that by choosing a suitable sign and amplitude for the quadratic optomechanical coupling (QOC), one can achieve enhanced and thermally robust stationary bipartite entanglement between the subsystems, improved mechanical ground-state cooling, and $Q$-quadrature squeezing of the mechanical mode beyond the 3-dB limit of squeezing. In particular, we find that in the presence of QOC with negative sign and in the resolved sideband regime the bipartite optical-mechanical entanglement can be increased by about 2 order of magnitude around the temperature of 1mK, and it can be preserved against thermal noise up to the ambient temperature of 0.1K. Furthermore, the QOC with positive sign can give rise to the enhancement of the mechanical ground-state cooling by about 1 order of magnitude in the optical and microwave red-detuned regime. We also find that for the positive sign of QOC and near the microwave resonance frequency the squeezing degree of the $Q$-quadrature of the mechanical mode can be amplified up to 15 dB. Such a hybrid electro-optomechanical system can serve as a promising platform to engineer an improved entangled source for quantum sensing as well as quantum information processing. 
\end{abstract}


\maketitle

\section{Introduction}
The optomechanical systems (OMSs) have recently attracted tremendous interest in both theoretical and experimental research in the field of quantum technologies, and have led to a deeper insight of quantum features at macroscopic scale \cite{aspelmeyer2014}. These systems, in which the quantized vibration of a mechanical oscillator (MO) is coupled to a quantized electromagnetic (EM) field via a radiation-pressure-type coupling \cite{bowen2015}, have been proposed to exploit in an increasingly diverse variety of applications, including cooling of mechanical resonators \cite{teufel2011,connell2010,chan2011}, generation of nonclassical states of the mechanical and optical modes \cite{hammerer2014,pontin2016,ockeloen2016,ockeloen20172,pirkkalainen2015}, realization of the optomechanically induced transparency (OMIT) \cite{xong2018,mikaeili2022,agarwal2010,weis2010,safavi2011,karuza2013,kronwald2013,motazedifard2022}, generating optomechanical entanglement  \cite{yang2019,palomaki2013,paternostro2007,dalafi2018,barzanjeh2019,mari2013},  ultraprecision quantum sensing and measurements \cite{kippenberg2007,tsang2010,buchmann2016,moler2017,liu2019,allahverdi2022,brunelli2019,ockeloen2018,
ockeloen20172,motazedifardAVS2021,ebrahimi2021,bemani2021}, quantum illumination or quantum radar \cite{barzanjehQIlluminationOMS2015,barzanjeh2020}, and detection and interferometry of gravitational waves \cite{courty2003}.

On the other hand, in the quantum communication networks it is necessary to convert the microwave frequency signal in the microprocessor to the optical frequency domain used in optical fibers and memories. To achieve an efficient frequency conversion, the microwave-optical entanglement is a vital component that should be engineered in the system. 
For this purpose, one can use an intermediate subsystem \cite{kurizki2015} such as ultra-cold atoms \cite{hafezi2022,verdu2009}, the ensemble of spins \cite{imamoglu2009,marcos2010}, and mechanical resonator \cite{safavi20112,regal2011,bochman2013,arnold2020}. 
In addition, an electro-optomechanical system could be a good candidate to achieve an optical-microwave converter \cite{vitali2007}. In such a system a MO is coupled to both optical and microwave modes, so one can convert optical to microwave frequency and vice versa through the MO \cite{thompson2008,forsch2020,barzanjeh2022,bonaldi2023,eshaghi2022}. Such a system can be used for quantum-illumination target detection \cite{barzanjehQIlluminationOMS2015}.
For example, in Refs.~\cite{andrews2014,haghighi2018} the MO is used as a coherent and effective bi-directional converter between optical and microwave frequencies.

In general, the optomechanical couplings can be categorized into two distinct types: the linear optomechanical coupling (LOC), in which the coupling strength is proportional to the displacement of the MO, and the quadratic optomechanical coupling (QOC), where the cavity field is coupled parametrically to the square of the displacement of the MO \cite{karuza2012,khorasani2018}. The regime of LOC involves one-phonon processes, while the QOC regime implies two-phonon processes. The QOC has been experimentally realized in the setups of membrane-in-the-middle OMSs \cite{thompson2008,javich2008}, where a flexible dielectric membrane is accurately positioned at a node or antinode of the intracavity standing wave, and ultracold atomic ensembles localized within a Fabry-Perot cavity \cite{purdy2010}. In addition, the QOC can arise even at the single-photon level \cite{liao2014} for which circuit quantum electrodynamics analogues have been proposed \cite{kim2015}. A measurement-based method has also been proposed to obtain an effective QOC \cite{vanner2011}. The quadratically coupled OMSs can bring about a variety of interesting quantum and nonlinear phenomena such as two-photon OMIT \cite{huang2011,karuza2013} and amplification \cite{si2017,liu2017}, photon and phonon blockade \cite{liao2013,shi2018}, quantum phase transition \cite{lu2018}, and tunable slow light \cite{zhan2013}. These systems have also been proposed for realizing quantum nondemolition  measurement of individual quantum jumps for continuous monitoring of the phonon shot noise \cite{clerk2010}, achieving macroscopic tunneling \cite{buchmann2012}, robust cooling and squeezing of a MO \cite{nunnenkamp2010,asjad2014,gu2015,banerjee2023}, engineering the macroscopic nonclassical states \cite{jacobs2009,shi2013,tan2013,abdi2016}, and realizing a highly sensitive mass sensor \cite{liu2019}. Moreover, some recent theoretical studies have considered both LOC and QOC together in different optomechanical configurations \cite{zhang2018,chao2021,xuereb2013}.

Motivated by the above-mentioned investigations, in this paper we intend to analyze the effects of the QOC on the steady-state bipartite entanglements, mechanical ground-state cooling, and mechanical squeezing in an electro-optomechanical system. For this purpose, we consider a microwave-optomechanical setup, which was proposed in Ref. \cite{barzanjeh2011} for generating a long-lived entanglement between an optical and a microwave cavity mode by means of their common linear coupling with a mechanical resonator. In the system under consideration, we assume that the MO is linearly coupled to the microwave mode, while, differently from Ref. \cite{barzanjeh2011}, it is simultaneously coupled to the optical mode through both LOC and QOC. 

It is found that the presence of even a weak QOC can give rise to an enhanced bipartite entanglement between the subsystems. In particular, the steady-state degree of optical-mechanical entanglement can be amplified by about 2 order of magnitude around the temperature of 1mK. Moreover, our results reveal that the QOC can also make the enhanced optical-mechanical entanglement robust against the thermal noise, such that it survives up to 0.1K. Other interesting findings of the present work concern the role of QOC in enhancing the ground state cooling and quadrature squeezing of the MO. We show that the QOC not only gives rise to a more effective red-sideband cooling of the MO motion to its quantum ground state, but also provides it to start from near sub-Kelvin temperatures without the need of using cryostat. In addition, we find that in the presence of  QOC the standard squeezing limit of 3 dB can be beaten for the Q-quadrature of the MO so that it may be squeezed to high degrees (up to 15 dB).

The paper is organized as follows. In Sec.~\ref{sec2} we describe the physical model of the system under consideration, give the quantum Langevin equations, and analyze the dynamical evolution of the mechanical mean field and the optical multistability as well as the dynamics of small quantum fluctuations. In Sec.~\ref{sec3} we will investigate the steady-state bipartite entanglement for all the bipartite subsystems. In Secs.~\ref{sec4} and ~\ref{sec5}, we respectively investigate the effects of QOC between the MO and the optical modes on the mechanical ground-state cooling and squeezing. We present our conclusions in Sec. ~\ref{sec6}. Finally, the Appendix provides some details of the calculations that are not explicitly given in the main text.

\section{theoretical description of the system}\label{sec2}

\begin{figure}
	\includegraphics[width=8.6cm]{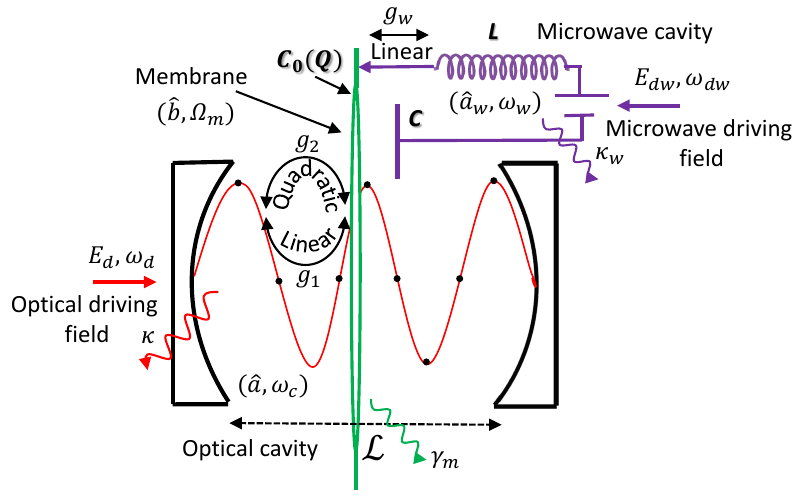}
	\caption{(colour online) Schematic illustration of the electro-optomechanical system under consideration. It consists of a single-mode optomechanical cavity with the membrane-in-the-middle geometry, where an oscillating membrane with frequency $\Omega_{m}$ and decay rate $\gamma_{m}$ couples both linearly and quadratically to the cavity field with frequency $\omega_{c}$ and decay rate $\kappa$. The membrane is also capacitively coupled to the microwave field mode of an LC circuit with frequency $\omega_{w}$ and decay rate $\kappa_{w}$. The optical cavity (microwave LC resonator) is driven by a classical field of amplitude $E_{d}$ ($E_{dw}$) with frequency $\omega_{d}$ ($\omega_{dw}$)}
	\label{fig1}
\end{figure}

The schematic of the system under consideration is depicted in Fig.~\ref{fig1}.The system consists of a single-mode Fabry-Perot-type optomechanical cavity with the membrane-in-the-middle configuration where the optical cavity field is coupled to a membrane which acting as a MO. The membrane can freely oscillate around its equilibrium position between node and anti-node of the optical mode inside the cavity, so that it couples to the cavity mode both linearly and quadratically \cite{zhang2018}. Simultaneously, the motion of the MO is capacitively coupled to the microwave field mode of an LC circuit which leads to a shift of capacitance, and thereby of the LC resonance frequency. Therefore, one obtains a linear interaction between the MO and the microwave mode. We assume that both the optical cavity and the microwave resonator are driven by classical fields. The Hamiltonian of the whole system reads

\begin{equation}
 \hat{H} =\hat{H}_{0} +\hat{H}_{int}+\hat{H}_{dr},\nonumber
\end{equation}
where
\begin{subequations}
    \begin{eqnarray}
&& \hat{H}_{0} = \hbar \omega_{c} \hat{a}^{\dagger} \hat{a} +\dfrac{\hbar \Omega_{m}}{2} (\hat{Q}^{2}+\hat{P}^{2})+\hbar \omega_{w} \hat{a}_{w}^{\dagger} \hat{a}_{w} ,\\
&& \hat{H}_{int} =-\hbar g_{1} \hat{a}^{\dagger} \hat{a} \hat{Q}-\hbar g_{2} \hat{a}^{\dagger} \hat{a} \hat{Q}^{2}- \hbar g_{w}\hat{a}_{w}^{\dagger} \hat{a}_{w} \hat{Q} ,\\
&& \hat{H}_{dr} =- i \hbar E_{d} (\hat{a} e^{i\omega_{d} t}-\hat{a}^{\dagger} e^{-i\omega_{d}t} )-i \hbar E_{dw} (\hat{a}_{w}e^{i\omega_{dw}t}-\hat{a}_{w}^{\dagger}e^{-i\omega_{dw}t}).\nonumber \\
  \label{hamiltonian}
  \end{eqnarray}
\end{subequations}

Here, $\hat{H}_{0}$ includes the free energies of the optical, mechanical, and microwave modes, with respective natural frequencies $\omega_{c}$, $\Omega_{m}$, and $\omega_{w}$. It is assumed that the membrane reflectivity is so low that the cavity field can be considered as a single-mode field. The optical and microwave modes are, respectively, described by the annihilation (creation) operators $\hat{a}$ ($\hat{a}^{\dagger}$) and $\hat{a}_{w}$ ($\hat{a}_{w}^{\dagger}$), such that $[\hat{a} ,\hat{a}^{\dagger}]=1$ and $[\hat{a}_{w} ,\hat{a}_{w}^{\dagger}]=1$. Furthermore, $\hat{Q}$ and $\hat{P}$ refer to the dimensionless displacement and momentum operators of the MO with the commutation relation $[\hat{Q},\hat{P}]=i$. The first two terms in $\hat{H}_{int}$ denote, respectively, the LOC and QOC between the MO and the optical field with respective single-photon linear and quadratic coupling strengths $g_{1}=(\omega_{c}/\mathcal{L})\sqrt{\hbar/m\Omega_{m}}$ and $|g_{2}|=\dfrac{8\pi^{2}c}{\lambda_{d}^{2}\mathcal{L}}\sqrt{\dfrac{R}{1-R}} \dfrac{\hbar}{m\Omega_{m}}$ ($m$ is the effective mass of the MO, $c$ is the speed of light, $\mathcal{L}$ denotes the length of the optical cavity, $\lambda_{d}$ is the wavelength of the optical driving field, and $R$ represents the reflectivity of the membrane) \cite{aspelmeyer2014,huang2011,bhattacharya2008}. Two remarks are in order at this point. First, in an optomechanical system with membrane-in-the-middle configuration, the coupling between the mechanical mode of the membrane and the cavity field depends on the equilibrium position of the membrane relative to the nodes of the cavity mode \cite{javich2008}. This causes a cavity detuning which is a periodic function of the membrane equilibrium position whose Taylor expansion up to the second order leads to the linear ($g_{1}$) and the quadratic ($g_{2}$) couplings of the membrane to the radiation pressure of the cavity. Second, if the membrane is located at local minima of the intracavity intensity, the quadratic coupling coefficient is positive. On the contrary, if the membrane is located at local maxima of the intracavity intensity, the quadratic coupling coefficient is negative \cite{seok2013,bhattacharya2008}. Therefore, both the absolute value of the ratio $g_{2}/g_{1}$ and its sign can be controlled and manipulated experimentally through the equilibrium position of the membrane relative to the nodes of the cavity mode \cite{thompson2008}.

 The third term in $H_{int}$  represents the linear coupling between the MO and the microwave mode with coupling strength $g_{w}=( \mu \omega_{w} /2d) \sqrt{h/m \Omega_{m}}$ in which $d$  is the unperturbed distance between the parallel plates of the capacitor, and $\mu =C_{0}(Q) /(C+C_{0}(Q))$ with $C_{0}(Q)=C_{0}d/(d-Q(t))$, where $C_{0}$ is the bare capacitance of the capacitor, and $C=C_{1}+C_{0}$ with $C_{1}$ being the stray capacitance in the microwave cavity (see Appendix ~\ref{appendix1}). The first (second) term in $H_{dr}$ describes the driving of the optical cavity (LC resonator) by a classical coherent field with the amplitude $E_{d}\,=\,\sqrt{2\,\mathcal{P}\,\kappa /( \hbar\,\omega_{d})}$ ($E_{dw}\,=\,\sqrt{2\,\mathcal{P}_{w}\,\kappa_{w} /( \hbar\,\omega_{dw})}$), the frequency $\omega_{d}$ ($\omega_{dw}$), and the power $\mathcal{P}$ ($\mathcal{P}_{w}$). Here $ \kappa $ ($ \kappa_w $) denotes the decay rate of the optical (microwave) mode.

By using the rotating reference frame defined by a unitary transformation $\hat{U}(t)=\exp (-i\omega_{d}\hat{a}^{\dagger} \hat{a} t-i \omega_{dw}\hat{a}_{w}^{\dagger}\hat{a}_{w}t)$, the total Hamiltonian of the system can be given by (see Appendix ~\ref{appendix1})
\begin{eqnarray}
  \label{eq2}
&&\!\!\!\!\!\!\!\!\!\! \hat{H}^{\prime}=\hbar \Delta_{0c}\hat{a}^{\dagger} \hat{a} +\dfrac{\hbar \Omega_{m}}{2}(\hat{Q}^{2}+\hat{P}^{2})+\hbar \Delta_{0w} \hat{a}_{w}^{\dagger}\hat{a}_{w} \nonumber\\
&& - \hbar g_{1}\hat{a}^{\dagger} \hat{a} \hat{Q} -\hbar g_{2} \hat{a}^{\dagger} \hat{a} \hat{Q}^{2} - \hbar g_{w} \hat{a}_{w}^{\dagger} \hat{a}_{w} \hat{Q}\nonumber\\
&&-i\hbar E_{d}(\hat{a}-\hat{a}^{\dagger})-i\hbar\,E_{dw}(\hat{a}_{w}-\hat{a}_{w}^{\dagger}),
\end{eqnarray}
where $\Delta_{0c} = \omega_{c} - \omega_{d}$ ($\Delta_{0w}= \omega_{w} -\omega_{dw}$) is the bare detuning of the optical (microwave) cavity from the optical (microwave) driving field. The dynamics of the electro-optomechanical system governed by the Hamiltonian $\hat{H}^{\prime}$ is fully determined by a set of nonlinear quantum Langevin equations in which the dissipation and fluctuation terms are taking into consideration, that is,
\begin{subequations}
\label{eq 3}
\begin{eqnarray}
&& \dot{\hat{a}} = -(i\Delta_{0c}+\kappa)\hat{a}+i (g_{1}\hat{Q}+g_{2}\hat{Q}^{2})\hat{a} +E_{d}+\sqrt{2\kappa} \hat{a}_{in},\\
&& \dot{\hat{Q}}=\Omega_{m}\hat{P},\\
&& \dot{\hat{P}}=-\Omega_{m}\hat{Q}+(g_{1}+2 g_{2}\hat{Q})\hat{a}^{\dagger}\hat{a}+ g_{w}\hat{a}_{w}^{\dagger}\hat{a}_{w} -\gamma_{m}\hat{P}\nonumber \\
&&\qquad +\sqrt{2\gamma_{m}}\hat{P}_{in},\\
&& \dot{\hat{a}}_{w} =- (i\Delta_{0w}+\kappa_{w})\hat{a}_{w}+i g_{w}\hat{a}_{w}\hat{Q}+E_{dw}+\sqrt{2\kappa_{w}} \hat{a}_{in,w},
  \end{eqnarray}
\end{subequations}
where $\gamma_{m}$  represents the dissipation rate of the mechanical mode. The system is affected by three uncorrelated quantum noise sources: the optical input vacuum noise $\hat{a}_{in}$ , the microwave input vacuum noise $\hat{a}_{in,w}$ , and the Brownian noise $\hat{P}_{in}$  acting on the MO. The quantum vacuum fluctuations $\hat{a}_{in}$  and $\hat{a}_{in,w}$, having zero mean values $<\hat{a}_{in}> =<\hat{a}_{in,w}>=0$, satisfy the Markovian correlation functions \cite{gardiner2000}
\begin{subequations}
\begin{eqnarray}
&& \!\!\!\!\! \langle \hat{a}_{in}(t)\hat{a}^{\dagger}_{in}(t^{\prime})\rangle =[n_{c}(\omega_{c})+1]\delta(t-t^{\prime}),\\
&& \!\!\!\!\! \langle \hat{a}^{\dagger}_{in}(t)\hat{a}_{in}(t^{\prime})\rangle =n_{c}(\omega_{c})\delta(t-t^{\prime}),\\
&& \!\!\!\!\! \langle \hat{a}_{in,w}(t)\hat{a}^{\dagger}_{in,w}(t^{\prime})\rangle =[n_{w}(\omega_{w})+1]\delta(t-t^{\prime}),\\
&& \!\!\!\!\! \langle \hat{a}^{\dagger}_{in,w}(t)\hat{a}_{in,w}(t^{\prime})\rangle =n_{w}(\omega_{w}) \delta(t-t^{\prime}),
\end{eqnarray}
\end{subequations}
in which $n_{c}(\omega_{c})\,=\dfrac{1}{e^{\hbar\,\omega_{c}\,/k_{B}\,T}-1}$ and $n_{w}(\omega_{w})\,=\dfrac{1}{e^{\hbar\,\omega_{w}\,/k_{B}\,T}-1}$ are, respectively, the mean thermal excitation numbers of optical and microwave fields, where $k_{B}$ is the Boltzmann constant and $T$ is the temperature of the surrounding environment. Since $\hbar\,\omega_{c}\,/k_{B}\,T\ll 1$ at optical frequencies, we can safely assume that $n_{c}(\omega_{c})\approx 0$, while $n_{w}(\omega_{w})$ cannot be neglected even the ambient temperature is quite low. In addition, in the limit of high mechanical quality factor, i.e., $\omega_{m}/\gamma_{m}\gg 1$, the zero-mean-value quantum Brownian noise $\hat{P}_{in}$  can be faithfully considered as a Markovian noise, with correlation function
\begin{equation}
 \langle \hat{P}_{in}(t)\hat{P}_{in}(t^{\prime})\rangle =[n_{m}(\Omega_{m})+1/2]\delta(t-t^{\prime}),
\end{equation}
where $n_{m}(\Omega_{m})\,=\dfrac{1}{e^{\hbar\,\Omega_{m}\,/k_{B}\,T}-1}$  is the mean thermal excitation number of the MO, like the optical and the microwave counterparts mentioned above.

We are interested in the mean-field dynamics of the system in a regime where the optical and microwave modes are strongly driven and the system is in the weak single-photon optomechanical coupling limit, i.e., $(g_{1}/\kappa) \sqrt{\hbar /m\Omega_{m}}, (g_{2}/\kappa)(\hbar/m\Omega_{m})\ll 1$.  Under these conditions,  the quantum Langevin equations of motion [Eqs.~\eqref{eq 3}] can be linearized by decomposing each Heisenberg operator $\hat{O}$ as a sum of its mean-field value $\langle \hat{O}\rangle$  and a quantum fluctuation operator $\delta \hat{O}$  with zero-mean value around the classical mean field, i.e., $\hat{O}=\langle \hat{O}\rangle+\delta \hat{O}$ with $\langle \delta\hat{O}^{\dagger} \delta\hat{O}\rangle / \langle \hat{O}^{\dagger} \hat{O} \rangle \ll 1$ \cite{aspelmeyer2014}. In addition, because of the quadratic coupling between the MO and the optical cavity field, we also need to calculate the steady-state solutions of the expectation values $\langle \hat{Q}^{2}\rangle$, $\langle \hat{P}^{2} \rangle$, and $\langle \hat{P} \hat{Q}+\hat{Q} \hat{P} \rangle$ \cite{sainadh2015}. The equations of motion for these quantities can be obtained by using Eqs.~\eqref{eq 3} together with the factorization assumption $\langle \hat{O}_{1} \hat{O}_{2} \hat{O}_{3} \rangle = \langle \hat{O}_{1} \rangle \langle \hat{O}_{2} \rangle \langle \hat{O}_{3}\rangle$, which is applicable in the weak single-photon coupling regime. By employing the operator decompositions
\begin{subequations}
\label{eq 4}
\begin{align}
\hat{a}\longrightarrow\,&\langle\,a\,\rangle\,+\delta\,\hat{a},\,\\
\hat{Q}\longrightarrow\,&\langle\,Q\,\rangle\,+\delta\,\hat{Q},\,\\
\hat{P}\longrightarrow\,&\langle\,P\,\rangle\,+\delta\,\hat{P},\,\\
\hat{a}_{w}\longrightarrow\,&\langle\,a_{w}\,\rangle\,+\delta\,\hat{a}_{w},\,\\
\hat{Q}^{2}\,\longrightarrow\,&\langle\,Q^{2}\,\rangle\,+\delta\,\hat{Q^{2}},\,\\
\hat{P}^{2}\,\longrightarrow\,&\langle\,P^{2}\,\rangle\,+\delta\,\hat{P^{2}},\,\\
\hat{P}\hat{Q}+\hat{Q}\,\hat{P}\,\longrightarrow\,&\langle\,PQ+QP\,\rangle\,+\delta(\hat{P}\hat{Q}+\hat{Q}\,\hat{P}),\,
\end{align}
\end{subequations}
the underlying equations of motion for the expectation values are obtained as
\begin{subequations}
\label{eq 7}
\begin{eqnarray}
&& \dot{\langle a \rangle}=-[i\Delta_{0c}+\kappa-ig_{1}\langle Q\rangle -ig_{2}\langle Q^{2}\rangle]\langle a \rangle +E_{d},\label{7a}\\
&& \dot{\langle Q \rangle}= \Omega_{m} \langle P \rangle ,\label{7b}\\
&& \dot{\langle P \rangle}=-(\Omega_{m}-2g_{2}\vert \langle a\rangle \vert^{2}) \langle Q \rangle +g_{1} \vert \langle a\rangle \vert ^{2}+g_{w} \vert \langle a_{w} \rangle \vert ^{2}-\gamma_{m} \langle P \rangle ,\label{7c}\\
&& \dot{\langle Q^{2} \rangle}=\Omega_{m} \langle PQ+QP\rangle ,\label{7d}\\
&& \dot{\langle P^{2} \rangle}=-(\Omega_{m}-2g_{2}\vert \langle a\rangle \vert^{2}) \langle PQ+QP\rangle +2 (g_{1} \vert \langle a \rangle \vert ^{2}+g_{w} \vert \langle a\rangle \vert^{2}) \langle P\rangle \nonumber \\
&&\hspace{1cm} -2\gamma_{m}\langle P^{2} \rangle +2\gamma_{m}(1+2n_{m}) ,\label{7e}\\
&& \dfrac{d}{dt}\langle PQ+QP\rangle =-2(\Omega_{m}-2g_{2} \vert \langle a\rangle \vert^{2})\langle Q^{2} \rangle +2\Omega_{m} \langle P^{2} \rangle \nonumber \\
&&\hspace{2.1cm} +2(g_{1} \vert \langle a \rangle\vert^{2}+g_{w}\vert \langle a_{w} \rangle \vert^{2}) \langle Q \rangle  -\gamma_{m} \langle PQ+QP\rangle ,\label{7f}\\
&&\!\!\!\!\! \langle \dot{a}_{w} \rangle =-(i \Delta_{0w}+\kappa_{w}-ig_{w}\langle Q \rangle)\langle a_{w} \rangle +E_{dw}\label{7g}.
\end{eqnarray}
\end{subequations} 

\subsection{Mean-field dynamics}
In order to investigate the semi-classical dynamics of the system, it is reasonable to use the adiabatic approximation. The reason for this arises from the fact that the damping rates of the optical and microwave modes are experimentally much faster than that of the mechanical mode, i.e.,  $\kappa\, , \kappa_{w}\gg \gamma_{m}$, so that the optical and microwave degrees of freedom follow the dynamics of the mechanical degree of freedom adiabatically. Therefore, from Eqs.~\eqref{7a} and \eqref{7g}, we obtain

\begin{figure}
\centering
\subfigure{\includegraphics[width=8.8cm]{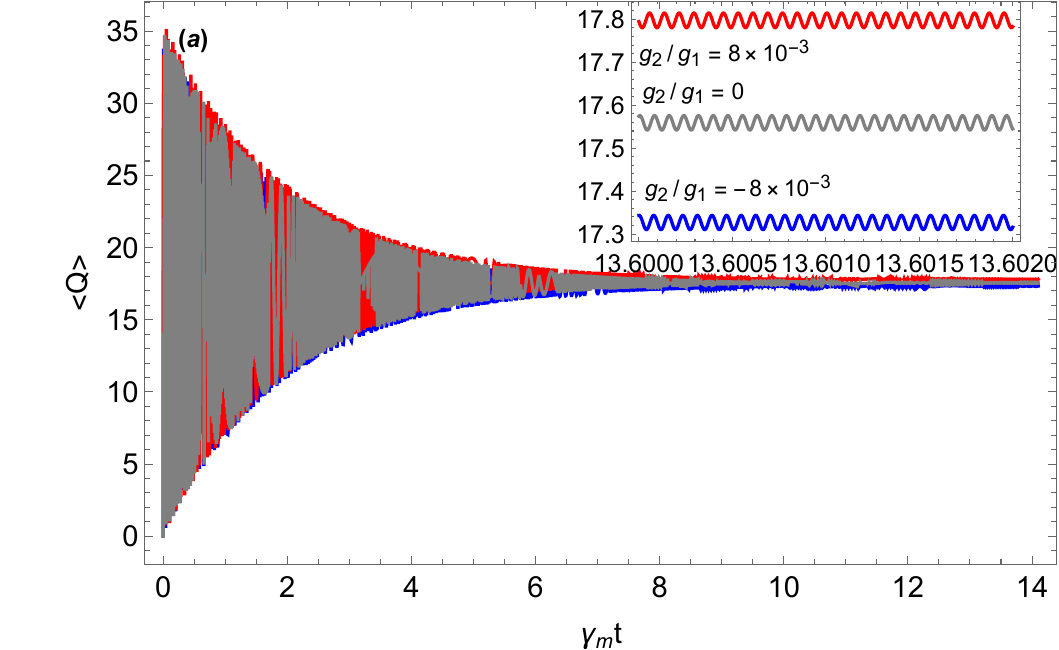}\label{fig2a}}
\subfigure{\includegraphics[width=8.8cm]{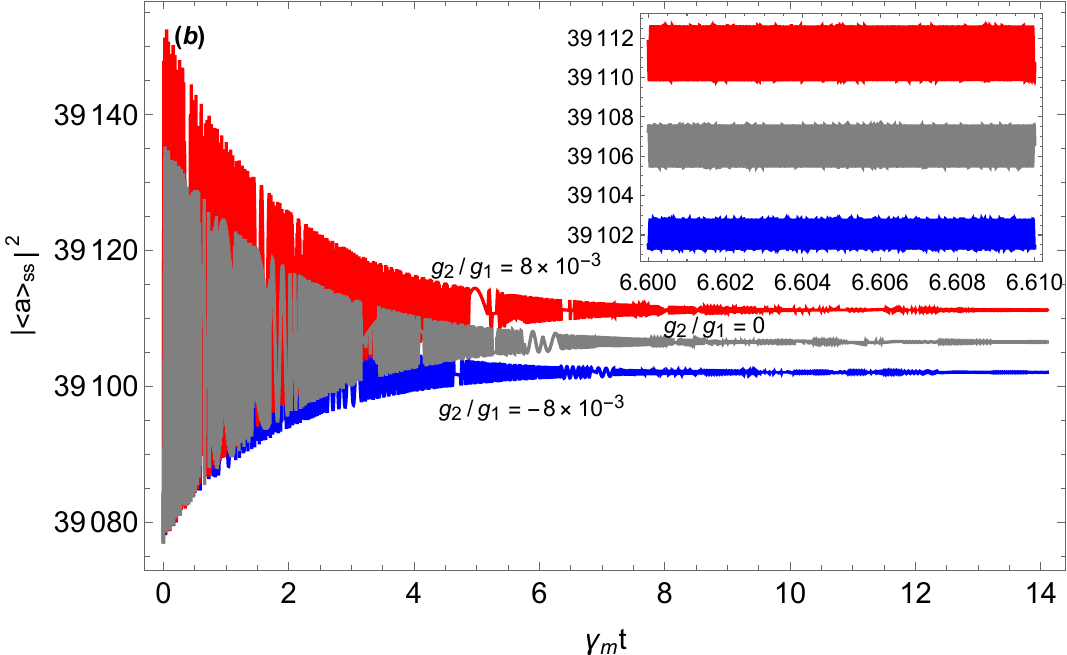}\label{fig2b}}

\caption{(colour online) (a) The mean value of the dimensionless $Q$-quadrature of the MO as a function of the normalized time $\gamma_{m} t$ for $g_{2}=8\times 10^{-3}$ (red), $g_{2}/g_{1}=0$ (grey), and $g_{2}/g_{1}=-8\times 10^{-3}$ (blue). (b) The mean number of optical cavity photons vs the normalized time $\gamma_{m}t$ for $g_{2}=8\times 10^{-3}$ (red), $g_{2}/g_{1}=0$ (grey), and $g_{2}/g_{1}=-8\times 10^{-3}$ (blue). The optical and microwave cavities are coherently driven by the input fields with powers $\mathcal{P} =30\mu$W and $\mathcal{P}_{w} =30\mu$W, respectively, and the bare detuning parameters have been fixed at $\Delta_{0c} =\Omega_{m}$ and $\Delta_{0w}= -\Omega_{m}$. The other parameters are $\lambda_{c} =810$nm, optical cavity length $\mathcal{L}=1$mm, $m=5$ng, $\Omega_{m}/2 \pi =10$MHz, $\mu =0.008$, separation between plates of capacitor $d=100$nm, microwave mode frequency $\omega_{w} /2 \pi =10$GHz, decay rate of the optical mode $\kappa =0.01 \Omega_{m}$, quality factor of mechanical mode $Q_m=8\times 10^{4}$, decay rate of the microwave mode $\kappa_{w} =0.005 \Omega_{m}$, and temperature $T=1$mK \cite{teufel20112}. The insets show the temporal behaviors of the same mean fields in a shorter time interval.}
\label{fig2}
\end{figure}

\begin{subequations}
  \label{eq 8}
 \begin{align}
 &\langle a \rangle_{ss} \approx \dfrac{E_{d}}{i\Delta_{c}+\kappa},\label{8a}\\
 &\langle\,a_{w}\,\rangle_{ss} \approx \dfrac{E_{dw}}{i\Delta_{w}+\kappa_{w}}\label{8b},
 \end{align}
\end{subequations}
where the effective detuning parameters $\Delta_{c}$ and $\Delta_{w}$ are defined as
\begin{subequations}
\begin{align}
&\Delta_{c}=\Delta_{0c}-\,g_{1}\,\langle\,Q\,\rangle_{ss}\,-g_{2}\langle\,Q^{2}\,\rangle_{ss}\,,\label{9a}\\
&\Delta_{w}=\Delta_{0w}-\,g_{w}\,\langle\,Q\,\rangle_{ss}\,\label{9b}.
\end{align}
\end{subequations}

Thus, by using Eqs.~\eqref{7b}-\eqref{7f} along with Eqs.~\eqref{8a} and \eqref{8b} we find a second-order non-linear differential equation for the  dimensionless mean field position of the MO as
\begin{equation}
\label{eq 10}
{\langle \ddot{Q} \rangle}+\gamma_{m} \langle \dot{Q} \rangle +\Omega_{m}^{\prime 2} \langle Q\rangle = \Omega_{m} (g_{1}\vert \langle a \rangle_{ss} \vert ^{2}+g_{w}\vert \langle a_{w} \rangle_{ss} \vert ^{2}),
\end{equation}
where $\Omega_{m}^{\prime} = \sqrt{\Omega_{m}(\Omega_{m}-2g_{2} \vert \langle a \rangle_{ss} \vert^{2})}$ denotes the effective frequency of the MO which explicitly depends on the mean value of intracavity photon number through the parameter $g_{2}$ of the QOC. 
In Fig.~\ref{fig2a} we have plotted the temporal evolution of the mean value of the $Q$ quadrature of the MO from the numerical solutions of Eqs.~\eqref{eq 10}, \eqref{8a}, and \eqref{8b} for different values of $g_{2}/g_{1}$. In this plot, we use the experimentally feasible parameters given in \cite{teufel20112}. We consider an optical cavity of length $\mathcal{L}=1$mm, natural frequency $\omega_{c}/2\pi=2.3\times 10^{15}$Hz ($ \lambda_c=818.564 \rm nm $), and damping rate $\kappa =0.01 \Omega_{m}$ which is coherently driven by a pump laser with wavelength $\lambda_{d}=810$nm and power $\mathcal{P} =30\mu$W. The membrane with mass $m=5$ng, damping rate $\gamma_{m}/2 \pi=125$Hz, and quality factor Q$_{m}=8\times 10^{4}$ oscillates with frequency $\Omega_{m}/2 \pi =10$MHz. The microwave resonator with natural frequency  $\omega_{w} /2 \pi =10$GHz and damping rate $\kappa_{w} =0.005 \Omega_{m}$ is driven by a microwave source with power $\mathcal{P}_{w} =30\mu$W. The other parameters are $\mu =0.008$ and the (unperturbed) separation between the plates of the capacitor $d=100$nm. As is seen from Fig.~\ref{fig2a}, the effect of the QOC on both the amplitude and  frequency of oscillations of the MO is negligibly small. This is mainly due to the fact that we work in the resolved sideband limit, i.e., $\kappa\ll \Omega_{m}$, so that $\Omega_{m}^{\prime}\cong \Omega_{m}$. Moreover, the presence of QOC does not affect the relaxation time of oscillations of the MO compared to that in its absence. In Fig.~\ref{fig2b} we have demonstrated the temporal behavior of the  optical mean field for three different values of QOC parameter. As can be seen, the overall effect of the QOC with a positive (negative) sign appears to be a little bit increase (decrease) in the mean number of cavity photons.

\subsection{Optical multistability}
We now examine the effect of the QOC on the optical multistability behavior of the system. To this end, we need to determine the steady-state solutions of the mean-field equations ~\eqref{7a}-\eqref{7g}. In the steady state, the mean fields of the optical and microwave modes are, respectively, given by Eqs.~\eqref{8a} and Eqs.~\eqref{8b}, while those of the mechanical mode read as 
\begin{subequations}
	\label{eq 6}
	\begin{align}
&\langle\,Q\,\rangle_{ss}=\dfrac{g_{1}\,\vert \langle\,a\,\rangle_{ss}\vert ^{2}+g_{w}\vert \langle\,a_{w}\,\rangle_{ss} \vert ^{2}}{\tilde{\Omega}_{m}},\label{11a}\\
&\langle\,P\,\rangle_{ss}\,=0,\label{11b}\\
&\langle\,Q^{2}\,\rangle\,_{ss}=\dfrac{\Omega_{m}(1+2n_{m})}{\tilde{\Omega}_{m}}+\langle\,Q\,\rangle_{ss}^{2},\label{11c}\\
&\langle\,P^{2}\,\rangle_{ss}=1+2\,n_{m},\label{11d}\\
&\langle\,PQ+QP\,\rangle_{ss}=0,\label{11e}
	\end{align}   
\end{subequations}
where $\tilde{\Omega}_{m}=\Omega_{m}-2g_{2}\,\vert \langle a \rangle_{ss} \vert ^{2}$. Solving this set of nonlinear algebraic equations gives us the steady-state value of the optical mode as a function of the detuning $\Delta_{0c}$ for different values of the QOC parameter $g_{2}$.

\begin{figure}
\centering
\subfigure{\label{fig3a}\includegraphics[width=8.6cm]{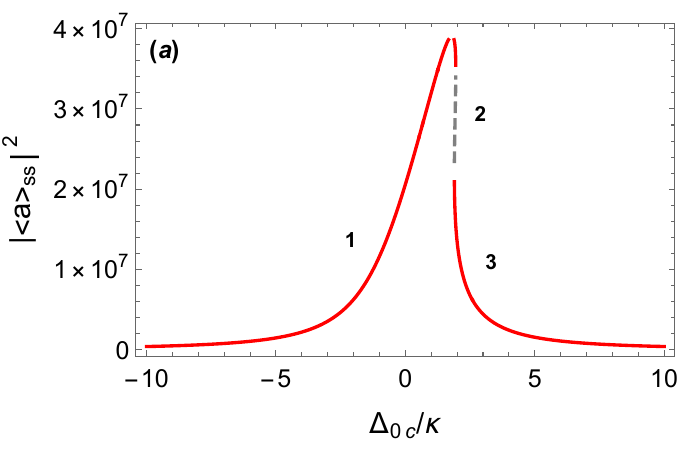}}

\subfigure{\label{fig3b}\includegraphics[width=8.6cm]{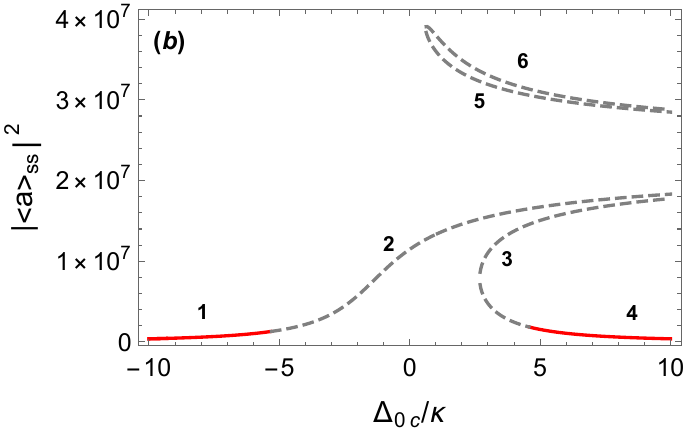}}

\subfigure{\label{fig3c}\includegraphics[width=8.6cm]{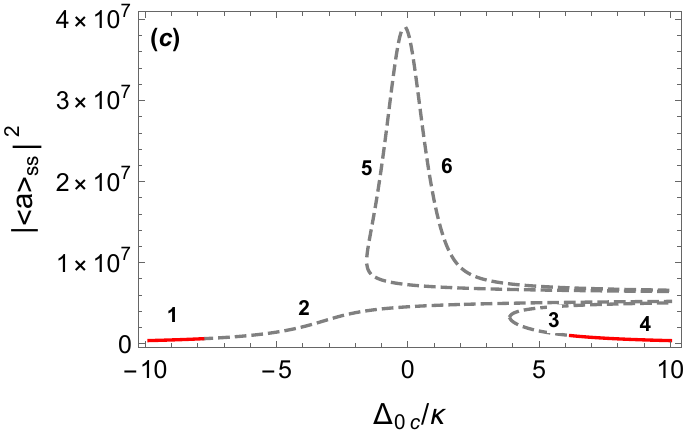}}
  \caption{(colour online) The steady-state mean value of the optical field in the cavity vs the dimensionless optical bare detuning $\Delta_{0c}/\kappa$ for different values of $g_{2}/g_{1}$: (a) $g_{2}=0$, (b) $g_{2}/g_{1}=10^{-3}$, and (c) $g_{2}/g_{1}=4\times 10^{-3}$. The red solid and gray dashed lines color correspond to stable and unstable branches, respectively. Here we have set $\Delta_{0w}=\Omega_{m}$, $\mathcal{P} =3\mu$W, and $\mathcal{P}_{w} =3\mu$W . The other parameters are the same as in Fig.~\ref{fig2}.}
   \label{fig3}
\end{figure}

In Fig.~\ref{fig3} we have plotted the steady-state mean number of photons in the optical cavity against the normalized detuning $\Delta_{0c}/\kappa$ in the absence of QOC, i.e., $g_{2}/g_{1}=0$ [Fig.~\ref{fig3a}] and in its presence with $g_{2}/g_{1}=10^{-3}$  [Fig.~\ref{fig3b}] and $g_{2}/g_{1}=4\times 10^{-3}$ [Fig.~\ref{fig3c}]. As is seen from Fig.~\ref{fig3}, in the absence of the QOC the system shows an ordinary optical bistability behavior for $\Delta_{0c}>2\kappa$. Here, the optical field is stable along the branches 1 and 3 while it is unstable along the branch 2. The stability conditions can be obtained by using the Routh-Hurwitz criterion \cite{dejesus1987} which imposes certain constraints on the system parameters (see the next subsection and Appendix~\ref{appendix3}  for the explanation thereof). Figures \ref{fig3b} and \ref{fig3c} exhibit the impact of QOC on the multistability behavior of the system. For the positive values of the QOC parameter we see multistability but when we check the Routh-Hurwitz stability condition only branches 1 and 4 are stable in the system and branches 2, 3, 5 and 6 are unstable for both $g_{2}=10^{-3}$ and $g_{2}=4\times 10^{-3}$. For negative values of the parameter QOC multistability will not occur.
\begin{figure}[h]
  \includegraphics[width=8.6cm]{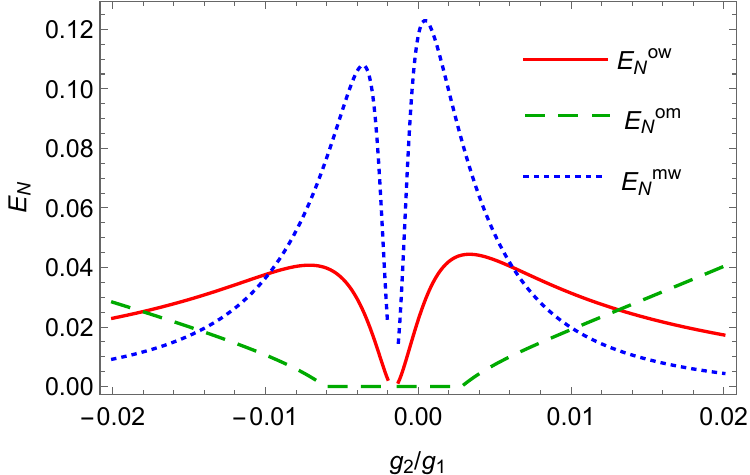}
   \caption{(Colour online) Bipartite logarithmic negativities $E_{N}^{ow}$, $E_{N}^{om}$, and $E_{N}^{mw}$ versus the ratio $g_{2}/g_{1}$. The red-solid, green-dashed and blue-dotted lines correspond, respectability, to the optical-microwave, optomechanical, and microwave-mechanic bipartite entanglements. Other parameters are the same as in Fig.~\ref{fig2}.}
  \label{fig4}
\end{figure}
\subsection{Stationary quantum fluctuations}
Having discussed the mean-field solutions, we now turn our attention to examine the dynamics of quantum fluctuations around the semiclassical steady state. The linearized quantum Langevin equations for the fluctuation operators can be expressed in the compact matrix form (see Appendix~\ref{appendix2})
\begin{equation}
\label{eq 9}
\delta \dot{\hat u}(t)\,=\,\textbf{A}\,\delta \hat u(t)\,+\,\delta \hat n\,(t)\,,
\end{equation}
where $\delta{\hat u}(t)$ (given by Eq.~\eqref{u}) is the vector of continuous-variable fluctuations and $\textbf{A}$ (given by Eq.~\eqref{A}) is the $6\times 6$ drift matrix. Furthermore, $\delta \hat n\,(t)$ (given by Eq.~\eqref{n}) represents the corresponding vector of noises.

The stationary properties of the quantum fluctuations can be examined by considering the steady-state condition governed by Eq.~\eqref{eq 9}. The steady state associated with this equation is reached when the system is stable, which occurs if and only if the real part of all the eigenvalues of the drift matrix $\textbf{A}$ are negative. These stability conditions can be deduced by applying the Routh-Hurwitz criterion which results in six independent inequalities, the explicit expressions of which are given in Appendix~\ref{appendix3}. Because of the linearized dynamics of the quantum fluctuations and since all noises are Gaussian, the steady state is a zero-mean Gaussian state which is fully characterized by the 6 ×6 stationary correlation matrix (CM) V, with components $V_{i\,j}=\langle \delta u_{i}(\infty )\delta u_{j}(\infty ) +\delta u_{j}(\infty ) \delta u_{i}(\infty ) \rangle /2$. By solving Eq.~\eqref{eq 9} we obtain
\begin{equation}
\label{eq 90}
\delta u(t)=M(t)\, \delta u(0)+\int_{0}^{\infty} dt^{\prime} M(t^{\prime})\, \delta n(t-t^{\prime}).
\end{equation}
where $M(t)=\exp(\textbf{A} t)$. Under the conditions that the system is stable then $M(\infty )=0$, and thus the steady-state solution reads as
\begin{equation}
\delta u_{i}(\infty)=\int_{0}^{\infty} dt^{\prime} \sum_{k} M_{i k}(t^{\prime})\,  \delta n_{k}(t-t^{\prime}).
\end{equation}
Therefore, the steady-state values of the CM matrix elements are given by
\begin{equation}
  \label{eq 100}
  V_{i j}=\sum_{k,l} \int_{0}^{\infty} d t \int_{0}^{\infty} dt^{\prime} M_{i k}(t) \, M_{j l}(t^{\prime})\,\tilde{D}_{k l}(t-t^{\prime}),
\end{equation}

\begin{figure}
\centering
\subfigure{\label{fig5a}\includegraphics[width=8.6cm]{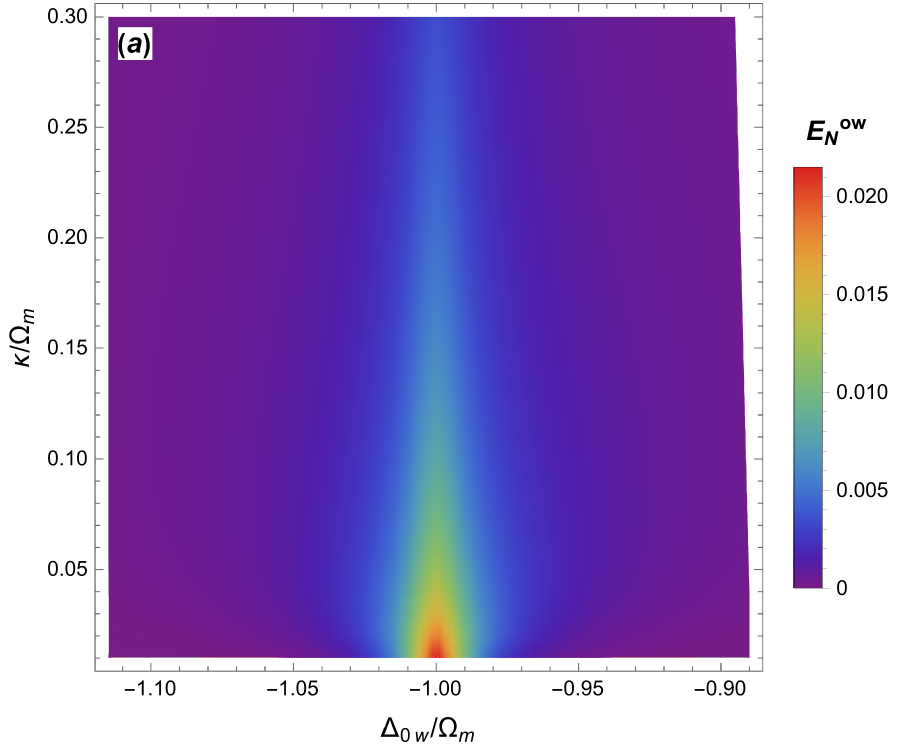}}

\subfigure{\label{fig5b}\includegraphics[width=8.6cm]{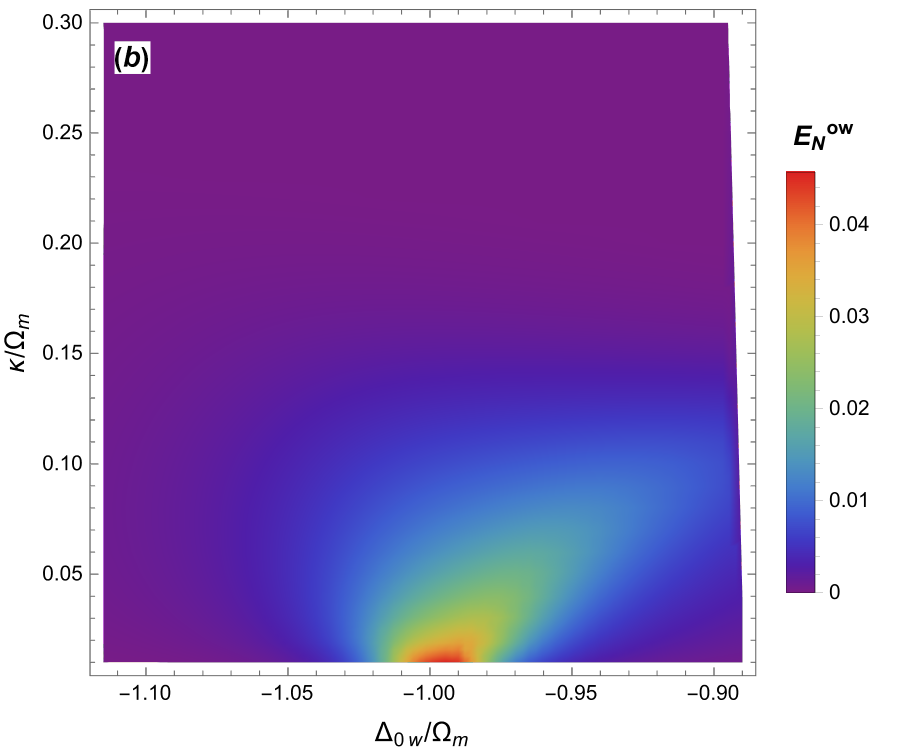}}

\subfigure{\label{fig5c}\includegraphics[width=8.6cm]{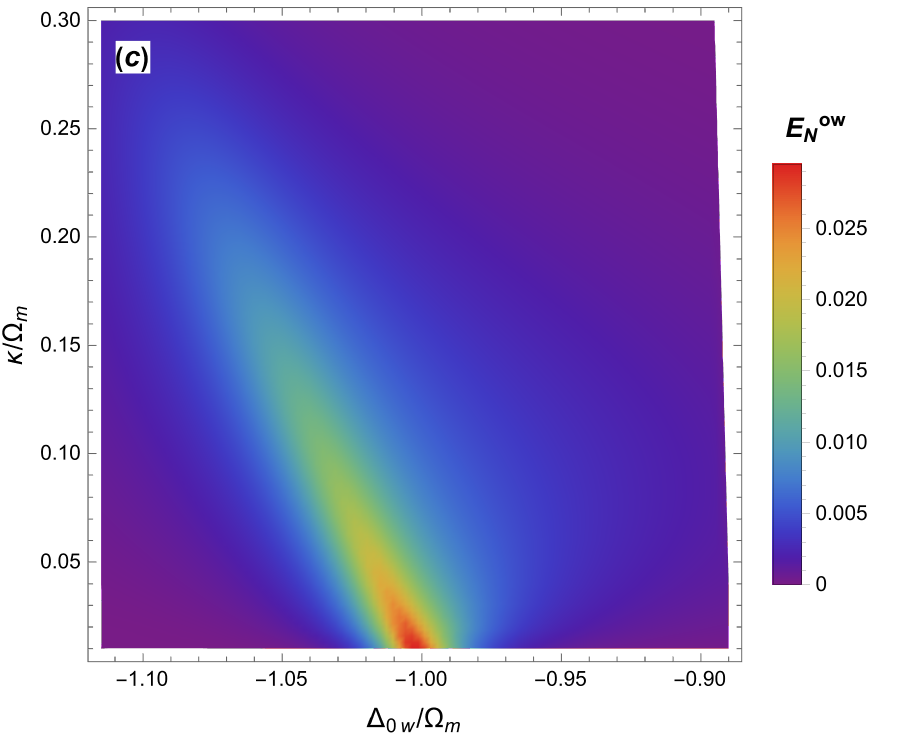}}
  \caption{(colour online) Density plots of the logarithmic negativity $ E_{N}^{ow} $ versus the dimensionless parameters $\Delta_{0w}/\Omega_{m}$ and $\kappa / \Omega_{m}$ for three different values of $g_{2}/g_{1}$: (a) $g_{2}=0$, (b) $g_{2}/g_{1}=4\times 10^{-3}$, and (c) $g_{2}/g_{1}=-4\times 10^{-3}$. Other parameters are the same as in Fig.~\ref{fig2}.}
   \label{fig5}
\end{figure}

where $\tilde{D}_{k l}(t-t^{\prime}) =\langle \delta n_{k}(t) \delta n_{l}(t^{\prime}) +\delta n_{l}(t^{\prime}) \delta n_{k}(t) \rangle /2$ is the matrix of stationary noise correlation function. Here, $\tilde{D}_{kl}(t-t^{\prime})=D_{kl}\delta(t-t^{\prime})$  in which $D=$ diag$[\kappa , \kappa ,0 ,\gamma_{m}(2 n_{m}+1) ,\kappa_{w}(2n_{w}+1) , \kappa_{w}(2n_{w}+1)]$  is a six-dimensional diagonal matrix.

The CM in the stationary state is thus obtained as
\begin{equation}
  \label{eq 11}
  V=\int_{0}^{\infty} dt^{\prime} M(t^{\prime})\, D\, M^{T}(t^{\prime}).
\end{equation}
If  the stability conditions are fulfilled, then according to the Lyapunov’s first theorem \cite{lyapunov1992}, Eq.~\eqref{eq 11} is equivalent to
\begin{equation}
  \label{eq 12}
  \textbf{A} V +V \textbf{A}^{T}= -D.
\end{equation}

\section{steady-state entanglement}\label{sec3}

In this section, we are interested in exploring the effect of QOC on the behavior of the bipartite continuous
variable entanglement between any pair in the tripartite system under consideration. As the CM $V$ is obtained by solving the Lyapunov equation [Eq.~\eqref{eq 12}], one can quantify the degree of steady-state bipartite entanglement by means of  the logarithmic negativity, $E_{N}$, which has been proved as an entanglement measure \cite{plenio2005,vidal2002} and can be defined as \cite{adesso2004}
\begin{equation}
   \label{eq 13}
   E_{N}=\max[0,-\ln 2\eta^{-}],
\end{equation}
where $\eta^{-}\equiv 2^{-1/2}[\Sigma(V_{bp})-\sqrt{\Sigma(V_{bp})^{2}-4 \det V_{bp}}]^{1/2}$ is the lowest symplectic eigenvalue of the partial transpose of the $4 \times 4$ CM, $V_{bp}$, associated with the selected bipartition, obtained by neglecting the rows and columns of the uninteresting modes

\begin{equation}
   \label{eq 14}
   V_{bp}=
   \begin{pmatrix}
   B & C \\
   C^{T} & B^{\prime}
   \end{pmatrix},
\end{equation}
and $\Sigma(V_{bp})=$det$B+$det$B^{\prime}-2$det$C$.

According to Eq.~\eqref{eq 13}, a Gaussian state is said to be entangled ($E_{N}>0$ ) if and only if $\eta^{-}<1/2$  which is equivalent to Simon’s necessary and sufficient entanglement criterion for certifying entanglement of two-mode system \cite{simon2000}.

In Fig.~\ref{fig4} we have plotted the three bipartite logarithmic negativities, $E_{N}^{ow}$ , $E_{N}^{om}$, and $E_{N}^{mw}$ (with the superscripts "$o$", "$m$", and "$w$" referring to the optical, mechanical, and microwave modes, respectively) versus the ratio $g_{2}/g_{1}$ . As is seen, the overall effect of the QOC manifests as the enhancement of the three kinds of bipartite entanglement, irrespective of its sign. However, the three logarithmic negativities do not behave in the same way, and the entanglement sharing is obvious. In particular, for $\vert g_{2}/g_{1}\vert \leq 4\times10^{-3}$ the microwave-mechanical entanglement, $E_{N}^{mw}$, increases significantly at the expense of the optomechanical entanglement, $E_{N}^{om}$,while the optical-microwave entanglement, $E_{N}^{ow}$, remains to exist.

\begin{figure}
\centering
\subfigure{\label{fig6a}\includegraphics[width=8.6cm]{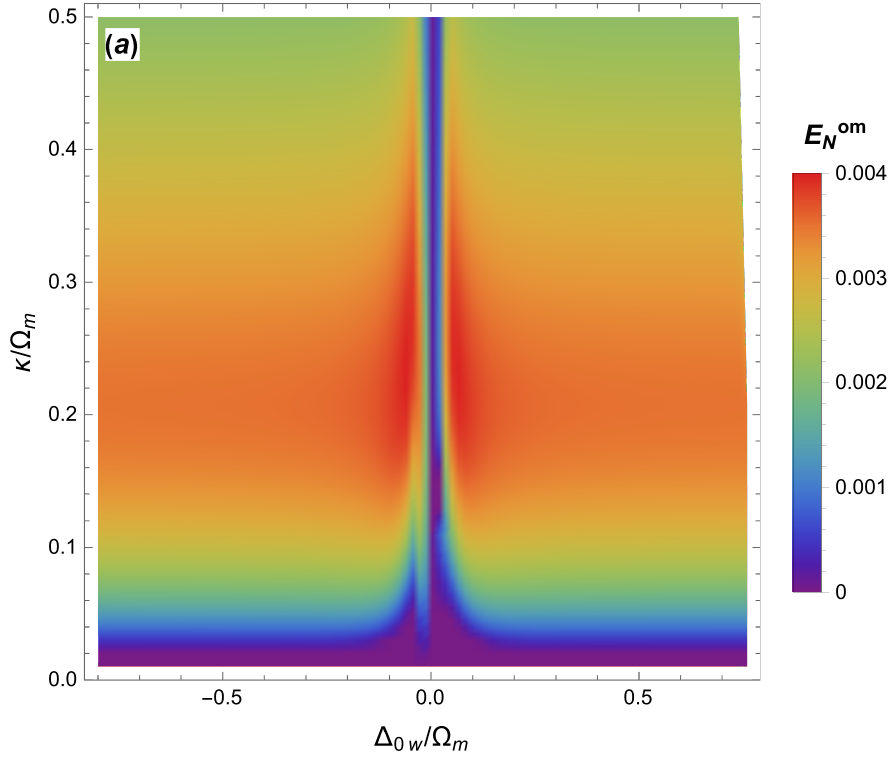}}

\subfigure{\label{fig6b}\includegraphics[width=8.6cm]{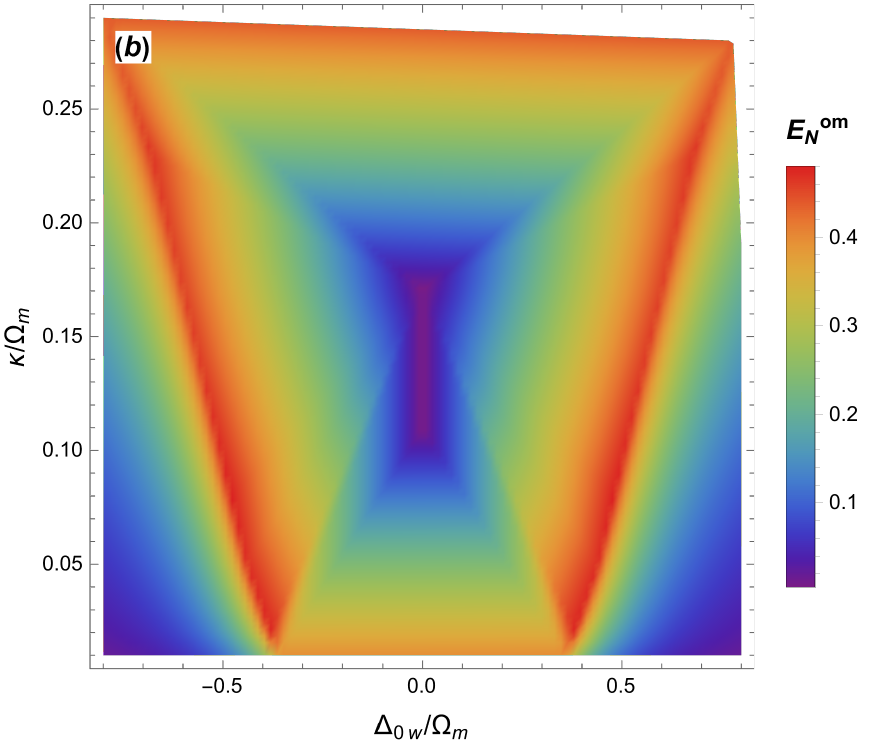}}

\subfigure{\label{fig6c}\includegraphics[width=8.6cm]{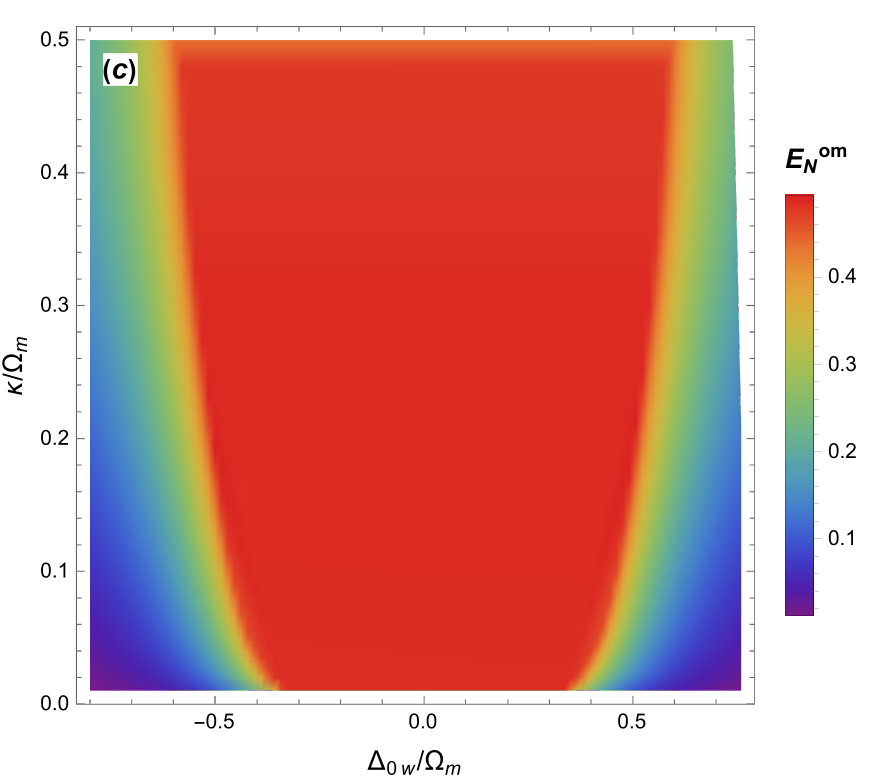}}
  \caption{(colour online)  Density plots of the logarithmic negativity $ E_{N}^{om} $ versus the dimensionless parameters $\Delta_{0w}/\Omega_{m}$ and $\kappa / \Omega_{m}$ for three different values of $g_{2}/g_{1}$: (a) $g_{2}=0$, (b) $g_{2}/g_{1}=4\times 10^{-3}$, and (c) $g_{2}/g_{1}=-4\times 10^{-3}$. Other parameters are the same as in Fig.~\ref{fig2}.}
   \label{fig6}
\end{figure}
\begin{figure}
\centering
\subfigure{\label{fig7a}\includegraphics[width=8.6cm]{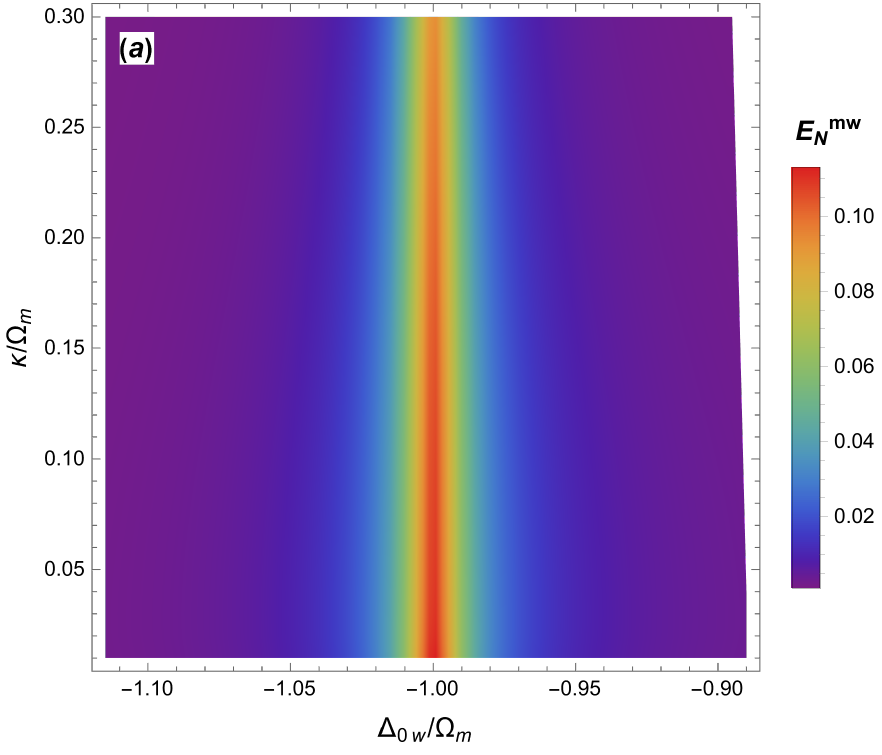}}

\subfigure{\label{fig7b}\includegraphics[width=8.6cm]{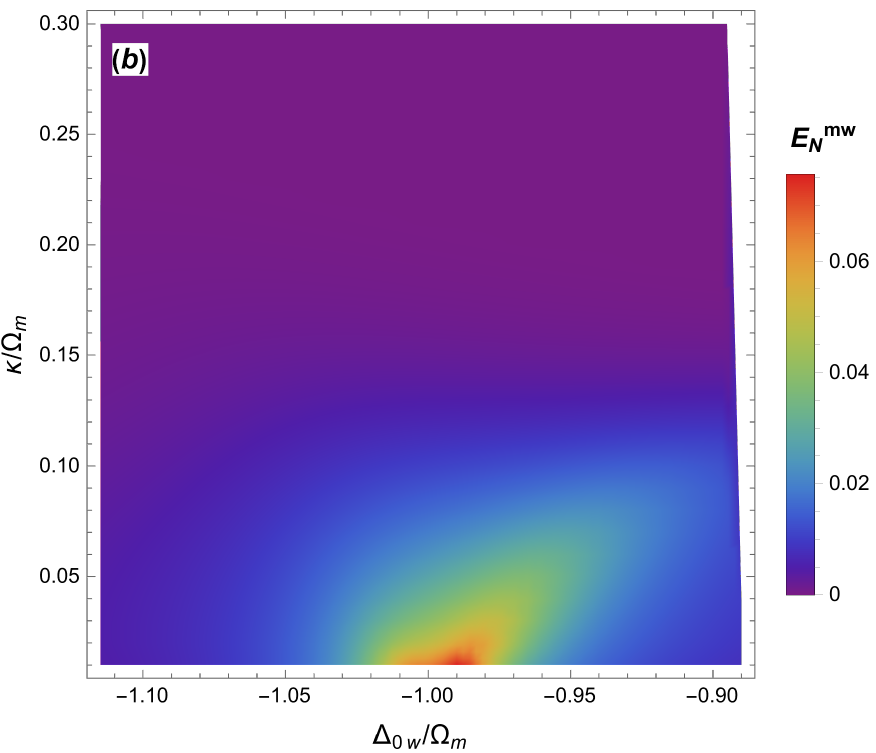}}

\subfigure{\label{fig7c}\includegraphics[width=8.6cm]{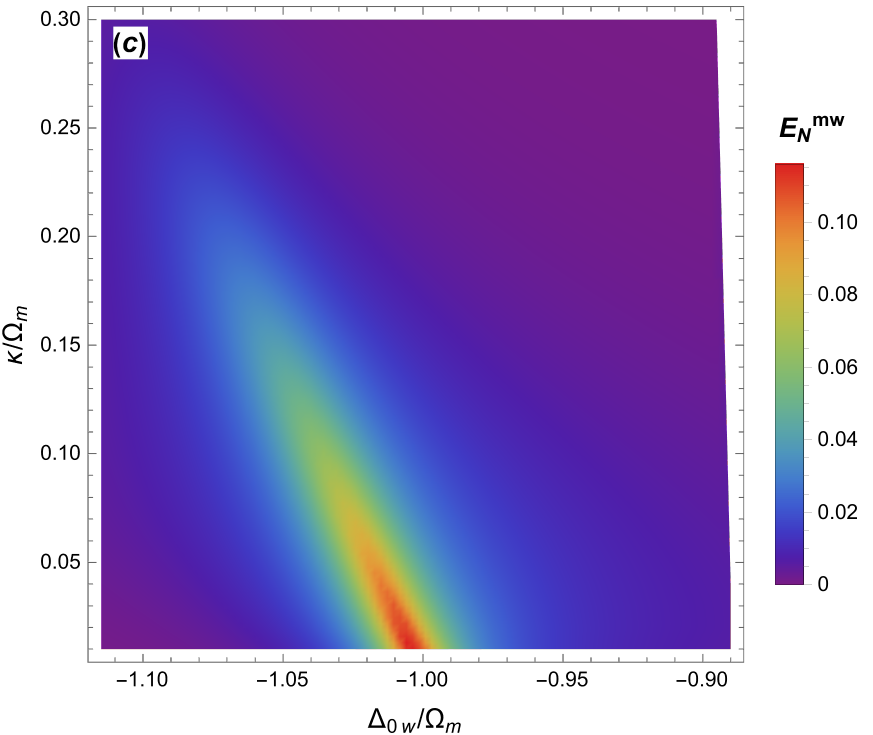}}
  \caption{(colour online)  Density plots of logarithmic negativity $ E_{N}^{mw} $ versus the dimensionless parameters $\Delta_{0w}/\Omega_{m}$ and $\kappa / \Omega_{m}$ for three different values of $g_{2}/g_{1}$: (a) $g_{2}=0$, (b) $g_{2}/g_{1}=4\times 10^{-3}$, and (c) $g_{2}/g_{1}=-4\times 10^{-3}$. Other parameters are the same as in Fig.~\ref{fig2}.}
   \label{fig7}
\end{figure}

The effects of QOC on the three kinds of bipartite entanglement are also illustrated in Figs.~\ref{fig5}-\ref{fig7}, where we have, respectively, plotted $E_{N}^{ow}$, $E_{N}^{om}$, and $E_{N}^{mw}$ versus $\Delta_{0w}/\Omega_{m}$ and $\kappa /\Omega_{m}$ for different values of the ratio $g_{2}/g_{1}$. Figures \ref{fig5a}-\ref{fig5c} show that in both cases of the absence of the QOC [Fig.~\ref{fig5a}] and the presence of it [Figs.\ref{fig5b} and \ref{fig5c}] the logarithmic negativity $E_{N}^{ow}$ is maximized around sideband $\Delta_{0w}=-\Omega_{m}$. However, in the presence of QOC with a positive sign the maximum amount of entanglement between the optical and microwave modes is about 0.04, i.e., about twice as large as for the cases without QOC and the QOC with a negative sign. Moreover, as expected, with increasing the optical cavity decay rate $\kappa$, the optical-microwave entanglement decreases drastically. We point out that the system under consideration may not be optimal for maximizing the entanglement between optical and microwave modes. Nevertheless, the present values of $E_{N}^{ow}$,even if not large, provides the possibility of achieving teleportation fidelities larger than those achievable with only classical means \cite{adesso2005,mari2008}. It is worth to remark here that it has been theoretically predicted and experimentally reported that higher stationary entanglement between microwave and optical fields (exceeding $10^{-1}$) can be achieved in some platforms, including  microwave cavity- optomagnomechanical systems \cite{fan2023}, directly coupled electro-optic systems \cite{rueda2019}, magnon-based hybrid systems \cite{zheng2024}, and optically pulsed superconducting electro-optical devices \cite{sahu2023}. According to Fig.~\ref{fig6a}, in the absence of QOC there is a weak entanglement between the optical and mechanical modes ($E_{N}^{om} \leq 0.004$) around $\Delta_{0w}=0$. On the Other hand, in the presence of the QOC, whether with positive sign [Fig.~\ref{fig6b}] or negative sign [Fig.~\ref{fig6c}], the optical-mechanical entanglement can be strengthened very intensively ($E_{N}^{om}\approx 0.5$). More interestingly, the comparison of Fig.~\ref{fig6c} with Figs.~\ref{fig6a} and \ref{fig6b} clearly reveals that in the presence of the QOC with negative sign the steady-state optomechanical entanglement with considerable robustness against the optical cavity loss is attainable over a relatively wide range of detuning $\Delta_{0w}$ in the resolved sideband regime ($\kappa\,<\Omega_{m}$). This behavior can be understood as follows. According to Eq.~\eqref{eq 10}, in the presence of quadratic coupling the mechanical spring constant is modified from $K=m\Omega_{m}^{2}$ to $K^{\prime}=m\Omega_{m}^{\prime2}=m\sqrt{\Omega_{m}(\Omega_{m}-2g_{2}\vert \langle a\rangle_{ss}\vert^{2})}$ which explicitly depends on the mean value of the intracavity photon number through the parameter $g_{2}$. This so-called optical spring effect can both soften or stiffen the mechanical oscillator depending on the sign of $g_{2}$. On the other hand, the stability of a cavity optomechanical system requires a balance between the radiation-pressure force, given by $F_{r}=\hbar \dfrac{\omega_{c}}{L}\vert \langle a \rangle_{ss} \vert^{2}$, and the mechanical restoring force. By using Eqs.\eqref{8a}, \eqref{8b}, \eqref{9a}, \eqref{9b}, \eqref{11a}, and \eqref{11c}, it is found that $F_{r}$ depend complicatedly on the system parameters $\kappa$, $\Omega_{m}$, $g_{2}/g_{1}$, and $\Delta_{0w}$. Since in the presence of QOC with a negative sign the mechanical restoring force increases, the radiation pressure force should also increases accordingly, leading to the enhancement of optomechanical entanglement. Numerical analysis shows that in this situation, the stability range as a function of detuning $\Delta_{0w}$ gets bigger in the resolved sideband regime ($\kappa / \Omega_{m}<1$). In other words, in this regime, the quadratic optomechanical coupling allows for optomechanical entanglement to persist against the cavity loss over a broader range of the detuning $\Delta_{0w}$. Beyond the good cavity limit ($\kappa / \Omega_{m}>1$) this situation is reversed; the larger the cavity damping rate is, the weaker the entanglement observed is. Concerning the mechanical-microwave entanglement, we find that the presence of QOC with a positive sign [Fig.~\ref{fig7b}] reduces while with a negative sign [Fig.~\ref{fig7c}] does not affect the maximum value of $E_{N}^{mw}$ around sideband $\Delta_{0w}=-\Omega_{m}$ compared to that in its absence [Fig.~\ref{fig7a}].

\begin{figure}
\centering
\subfigure{\label{fig8a}\includegraphics[width=8.6cm]{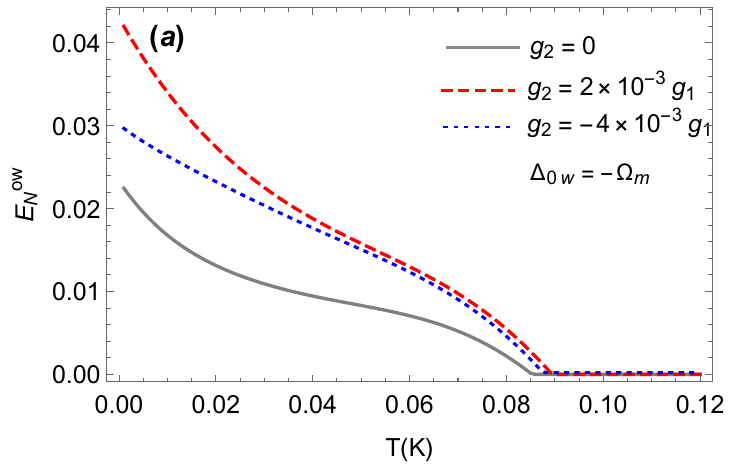}}

\subfigure{\label{fig8b}\includegraphics[width=8.6cm]{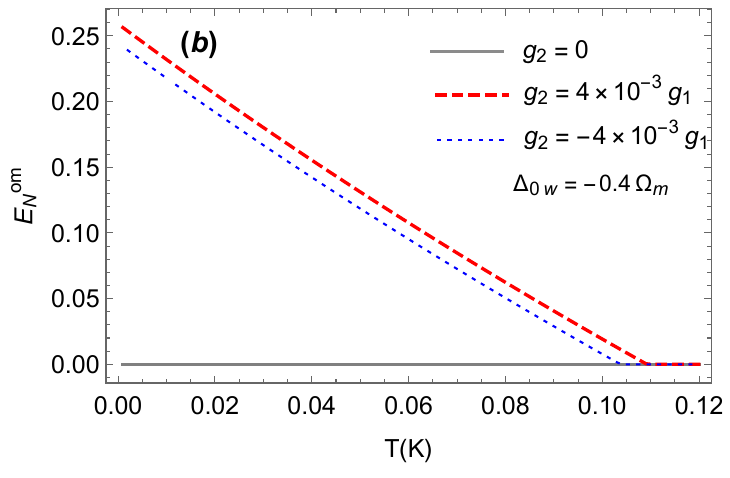}}

\subfigure{\label{fig8c}\includegraphics[width=8.6cm]{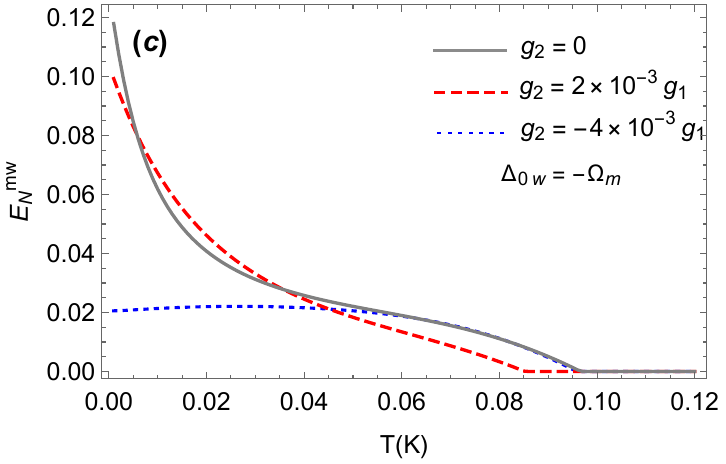}}
   \caption{(colour online) The logarithmic negativity $E_{N}$ between the (a) optical and microwave, (b) optical and mechanical, (c) mechanical-microwave modes versus environmental temperature $T$ for different values of $g_{2}/g_{1}$. In (a) and (c) we take $\Delta_{0w}=-\Omega_{m}$ and in (b) $\Delta_{0w}=-0.4\Omega_{m}$. Other parameters are the same as in Fig.~\ref{fig2}.}
   \label{fig8}
\end{figure}

\begin{figure}
   \includegraphics[width=8.6cm]{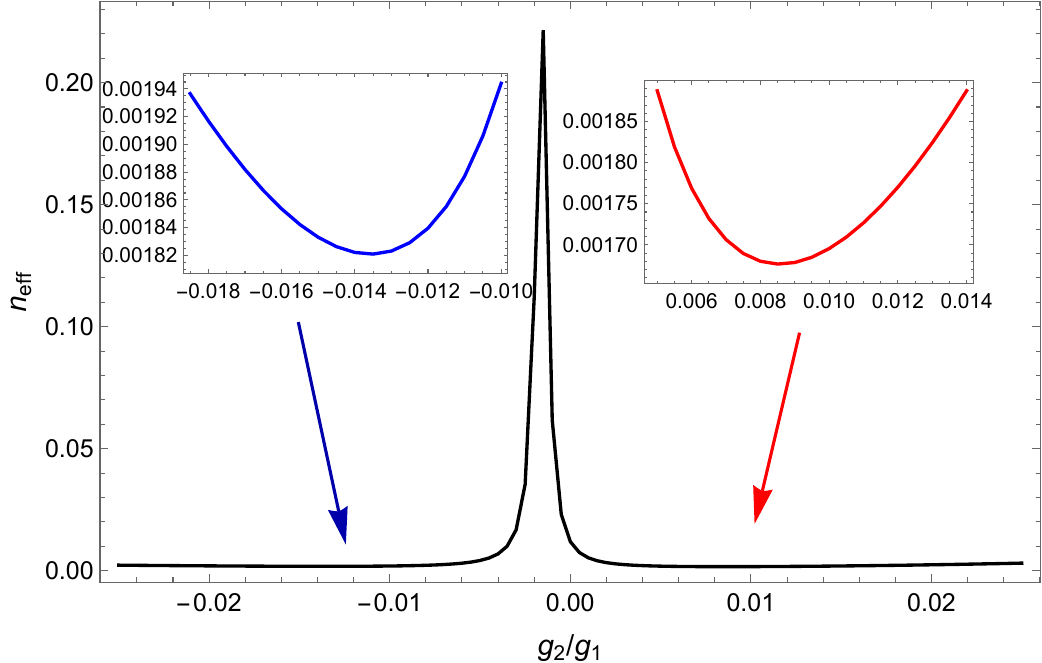}
   \caption{(color online) Effective mean phonon number of the mechanical mode $n_{\rm eff}$ vs $g_{2}/g_{1}$. The left and right insets show the enlarged views of the regions where $n_{\rm eff}$ reaches its minimum value for negative and positive values of QOC, respectively. Here we have set $\Delta_{0w} /\Omega_{m}=1$, and the other parameters are the same as in Fig.~\ref{fig2} }
   \label{fig9}
\end{figure}

\begin{figure*}
\includegraphics[width=14cm]{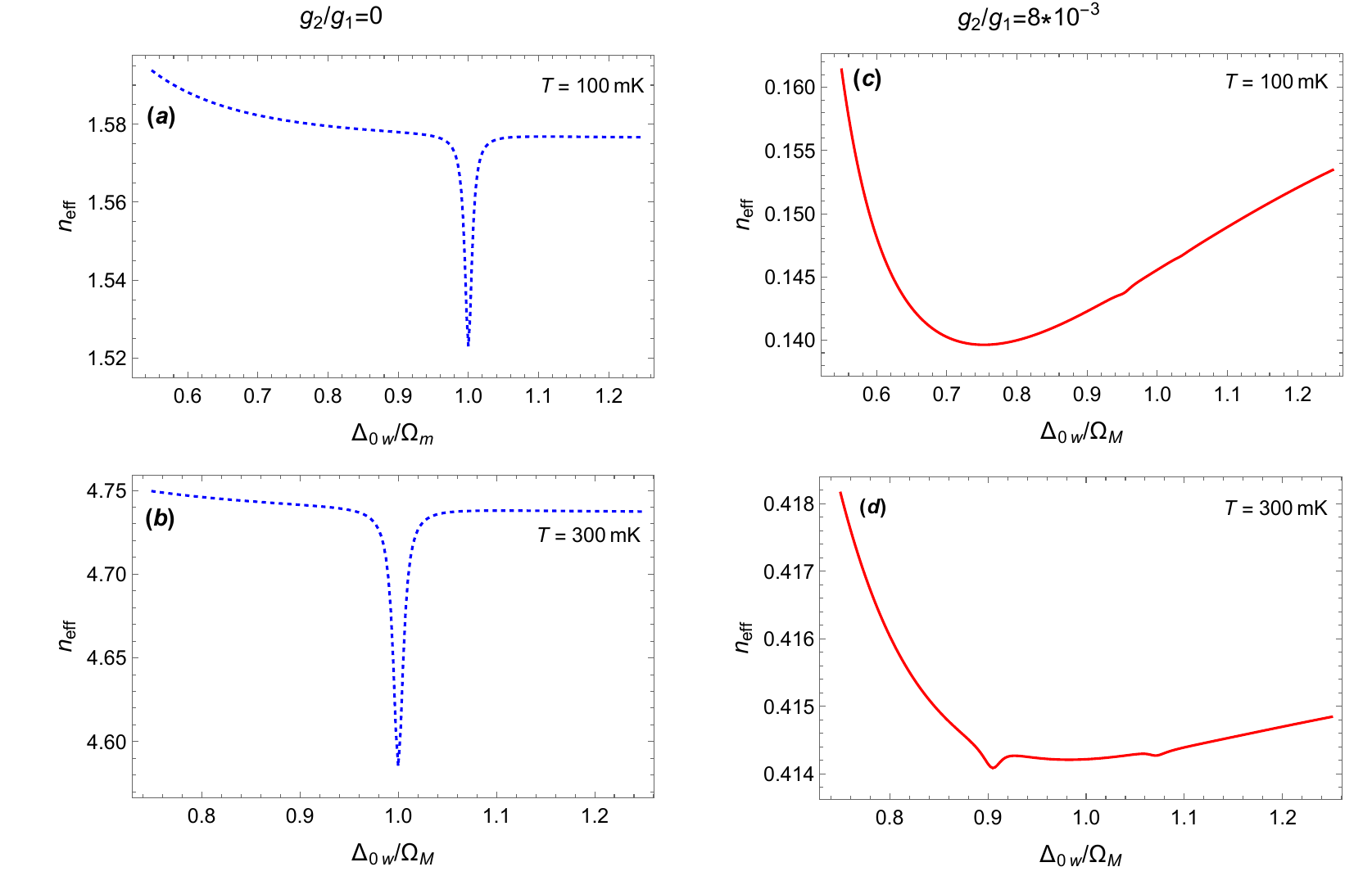}
	\caption{(color online) The effective mean phonon number, $n_{\rm eff}$, vs the dimensionless microwave detuning $\Delta_{0w} /\Omega_{m}$ for different temperatures of the environment. The left panels correspond to $g_{2}=0$, whereas the right ones correspond to $g_{2}=8\times 10^{-3}g_{1}$. All the other parameters are the same as in Fig.~\ref{fig2}}
	\label{fig10}
\end{figure*}

As is well known, systems unavoidably suffer from thermal noise, thereby, in general reducing the quantum correlations. The typical investigation in this context is to identify the temperature up to which the entanglement survives which measures its robustness against thermal fluctuations. Here, we are going to explore the influence of the QOC on the thermal robustness of the bipartite entanglements in the system under consideration. For this purpose, we have plotted in Figs.~\ref{fig8a}-\ref{fig8c} the dependence of bipartite logarithmic negativities, $E_{N}^{ow}$, $E_{N}^{om}$, and $E_{N}^{mw}$ on the environment temperature $T$ for different values of the QOC parameter. These figures clearly indicate that the behavior of the entangled subsystem pairs is sensitive to the environment temperature, such that the degree of bipartite entanglement decreases and eventually ceases to exist with increasing environmental temperature. Moreover, the figures show that the effect of QOC manifests as the change in the critical temperature $T_{c}$ which is defined as $T\geqslant T_{c}$, $E_{N}=0$. As is seen from Fig.~\ref{fig8a}, when the bare microwave detuning is set to $\Delta_{0w}=-\Omega_{m}$ (optimal microwave detuning value that maximizes $E_{N}^{ow}$ , see Fig.~\ref{fig5}) the optical-microwave entanglement $E_{N}^{ow}$ in the absence of QOC ($g_{2}=0$) survives until the critical temperature $T_{c}$ reaches about $0.084$K. On the other hand, in the presence of QOC, either with positive or negative sign, the optical-microwave entanglement $E_{N}^{ow}$ at temperatures below the critical temperature is stronger than the case in which $g_{2}=0$. However, as is seen, $E_{N}^{ow}$ disappears at almost the same critical temperature for both cases of $g_{2}=0$ and $g_{2}\neq 0$ .This means that when the temperature reaches the critical value, the effect of the QOC on $E_{N}^{ow}$ is nearly balanced with that of thermal noise. Figure \ref{fig8b} illustrates the behavior of the optical-mechanical entanglement with respect to temperature where we have set $\Delta_{0w}=-0.4\Omega_{m}$ (optimal microwave detuning value for the optical-mechanical entanglement in the presence of QOC, see Fig.~\ref{fig6}). As can be seen, there is no entanglement between the optical and mechanical modes when $g_{2}=0$. Nevertheless, the thermal robustness of the optical-mechanical entanglement is enhanced in the presence of QOC. It survives up to $T_{c}=0.110 $K and $T_{c}=0.105 $K for $g_{2}=4\times 10^{-3}g_{1}$  and $g_{2}=-4\times 10^{-3}$, respectively. This behavior is different from the one obtained for $E_{N}^{ow}$. That is to say, for the optical- mechanical $E_{N}^{om}$ , the QOC can effectively overcome the effect of thermal noise compared to Fig.~\ref{fig8a}. Lastly, we examine in Fig.~\ref{fig8c} the thermal robustness of the mechanical-microwave entanglement $E_{N}^{mw}$ at optimal microwave detuning $\Delta_{0w}=-\Omega_{m}$ (see Fig.~\ref{fig7}) for different values of the QOC parameter $g_{2}$. As can be seen, in contrast to the case of optical-mechanical entanglement, the presence of the QOC decreases the robustness of the entanglement between the mechanical and microwave modes compared to that in its absence.

\section{ground-state cooling of mechanical oscillator}\label{sec4}

In this section we are going to investigate the role of QOC on cooling the MO close to its quantum ground state. The steady-state mean energy of the MO is given by 
\begin{equation}
   \label{eq 17}
   U=\dfrac{\hbar \Omega_{m}}{2}( < \delta \hat{Q}^{2} > +<\delta \hat{P}^{2} >)=\hbar \Omega_{m}(n_{\rm eff}+\dfrac{1}{2}),
\end{equation}
 where $n_{\rm eff}$ is the mean phonon number of the MO corresponding to an effective temperature $T_{\rm eff}=\hbar \Omega_{m}/[k_{B}\ln(1+1/n_{\rm eff})]$ \cite{barzanjeh20112,shahidani2014}. The ground-state cooling of the MO is achieved whenever $n_{\rm eff}\simeq 0$ or $U\simeq \hbar \Omega_{m}/2$. This can be realized if $\langle \delta \hat{Q}^{2}\rangle=\langle \delta \hat{P}^{2} \rangle\simeq 1/2$ in the steady state. Once the CM $V$ is obtained, one can calculate $n_{\rm eff}$ from its diagonal elements as \cite{wang2023}
\begin{equation}
   \label{eq 18}
   n_{\rm eff}=\dfrac{V_{33}+V_{44}-1}{2}.
\end{equation}

In Fig.~\ref{fig9}, we have plotted the effective mean phonon number of the MO versus the ratio $g_{2}/g_{1}$ when $\Delta_{0w}=\Omega_{m}$. It is seen that in the presence of the QOC with positive sign, the effective mean number of phonons goes to a minimum of about $1.68\times 10^{-3}$ at $g_{2}/g_{1}=+8\times 10^{-3}$ [see the right inset of Fig.~\ref{fig9}] and the MO is cooled down to its ground state. On the other hand, when examining the negative region from right to left, we find that the mean number of phonons increases sharply as the absolute value$\mid g_{2}/g_{1}\mid$ increases, reaching the maximum value $0.24$ at $g_{2}/g_{1}=-2\times10^{-3}$, and after that it drops to a minimum of about $1.70\times 10^{-3}$ at $g_{2}/g_{1}=-0.013$ [see the left inset of Fig.~\ref{fig9}]. Hence, the positive values of QOC can provide a better cooling performance as compared to its negative values. In Fig.~\ref{fig10}, we have numerically plotted the effective mean phonon number of the MO as a function of the dimensionless detuning of the microwave cavity $\Delta_{0w}/\Omega_{m}$, for various bath temperatures of the MO. The left panels in the figure correspond to the case of absence of QOC ($g_{2}/g_{1}=0$), while the right panels correspond to its presence with $g_{2}/g_{1}=+8\times 10^{-3}$ for which the optimal ground state cooling of the MO is achieved [see Fig.~\ref{fig9}]. As is seen, in the absence of QOC [Figs.~\ref{fig10}(a) and ~\ref{fig10}(b)], when $\Delta_{0w}=\Omega_{m}$ the effective mean number of phonons of the mechanical mode goes to a minimum whose value increases with increasing the mechanical bath temperature $T$, as expected. On the other hand, we find that in the presence of QOC [Figs.~\ref{fig10}(c) -~\ref{fig10}(d)], the minimum attainable value of the effective number of phonons is reduced by 1 order of magnitude as compared to that for the case of absence of QOC. In fact, at each temperature of the mechanical bath there exists a certain range of $\Delta_{0w}/\Omega_{m}$, which we refer to it as a mechanical cooling window, over which the QOC gives rise to the reduction of $n_{\rm eff}$. As understood from Figs.~\ref{fig10}(c) and ~\ref{fig10}(d), at environment temperatures more than $T =0.1$K, the presence of QOC ensures that the MO is cooled down to its ground state. Interestingly, Fig.~\ref{fig10}(d) reveals that even at the sub-Kelvin temperature of $T =0.3$K the mean number of phonons is less than unity, which implies that the MO is near its ground state, while this is not the case in the absence of QOC.

 \begin{figure}
	\centering
\subfigure{\label{fig11a}\includegraphics[width=8.6cm]{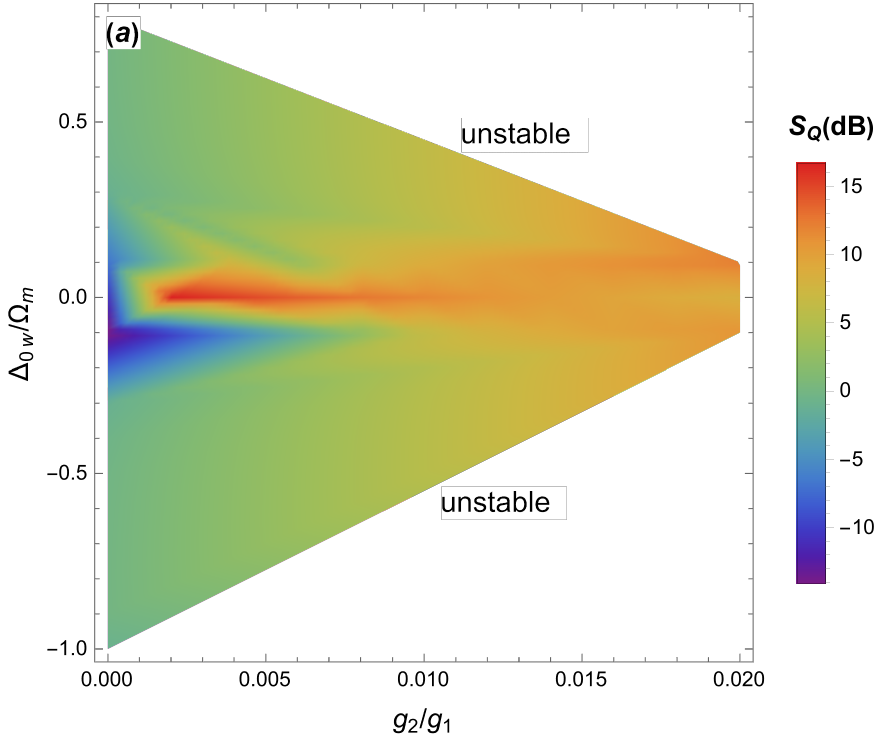}}

\subfigure{\label{fig11b}\includegraphics[width=8.6cm]{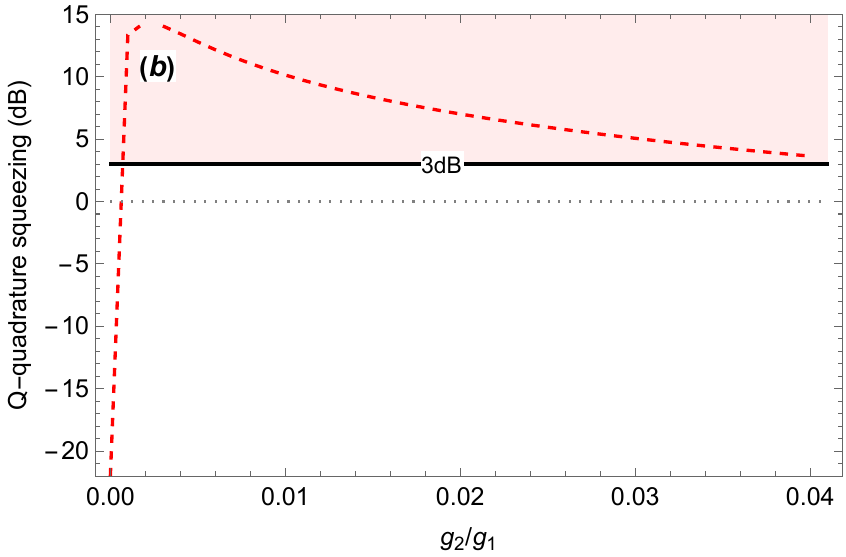}}
	\caption{(Color online)(a) Density plot of $S_{Q}$ vs $g_{2}/g_{1}$ and the normalized bare dimensionless detuning of the microwave cavity $\Delta_{0w}/\Omega_{m}$, and (b) plot of $S_{Q}$ vs $g_{2}/g_{1}$ for $\Delta_{0w}=0$. The shadowed pink region corresponds to squeezing beyond the 3 dB limit. Here we have set $\mathcal{P}=1 mW$ and $\mathcal{P}_{w}=1 mW$. The other parameters are the same as in Fig.~\ref{fig2}. }
	\label{fig11}
\end{figure}

It should be noted that the condition $n_{\rm eff}<1$ is not the only requirement for the ground-state cooling of the mechanical mode. In order to achieve the quantum ground-state cooling it is also necessary that both momentum and displacement variances tend to $\langle \delta \hat{P}^{2}\rangle \approx \langle \delta \hat{Q}^{2}\rangle \approx 1/2$ that is the energy equipartition should be satisfied in the optimal regime close to the ground state. Otherwise, the steady state of the system is not a thermal equilibrium state, and thus one cannot define an unambiguously effective temperature for the MO in this case \cite{genes20081}. We have examined this second requirement in Appendix~\ref{appendix4}.

\section{steady-state mechanical squeezing}\label{sec5}

In this section, we are going to investigate the effect of QOC on the mechanical quadrature squeezing. If $\hat{\delta Q}$ and $\hat{\delta P}$ are the dimensionless quadrature components of the MO which satisfy the commutation relation $[\hat{\delta Q}, \hat{\delta P}]=i$, the Heisenberg uncertainty relation is given by $\Delta Q\, \Delta P \geq \sigma_{\rm zpf}$ where $\sigma_{\rm zpf}=\vert \langle [\hat{\delta Q},\hat{\delta P}]\rangle \vert /2$ is the zero-point fluctuation and $\Delta Q:=\sqrt{\langle \hat{\delta Q}^{2} \rangle -\langle \hat{\delta Q} \rangle^{2}}$ and $\Delta P :=\sqrt{\langle \hat{\delta P}^{2}\rangle - \langle \hat{\delta P} \rangle^{2}}$ denote the quadrature uncertainties. If a quantum mechanical oscillator is in a state where one of the variances $\sigma_{j}=(\Delta j)^{2}$, for $j=Q$ or $P$, is less than $\sigma_{\rm zpf}=1/2$, then it can be judged that mechanical squeezing has been generated. The degree of the squeezing can be expressed in the decible (dB) unit, which can calculate by \cite{agarwal2016}
\begin{eqnarray}
&& S_j=-10 \log_{10}\left(\frac{\sigma_{j}}{\sigma_{\rm zpf}} \right).
\end{eqnarray}

\begin{figure}
   \centering
\subfigure{\label{fig12a}\includegraphics[width=8.6cm]{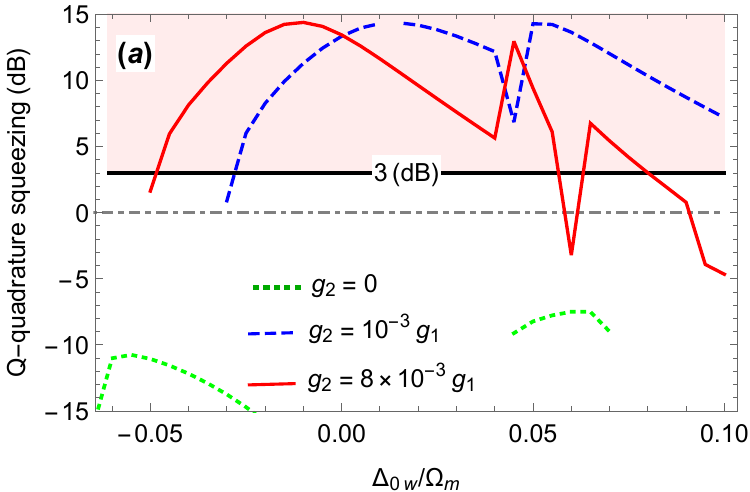}}

\subfigure{\label{fig12b}\includegraphics[width=8.6cm]{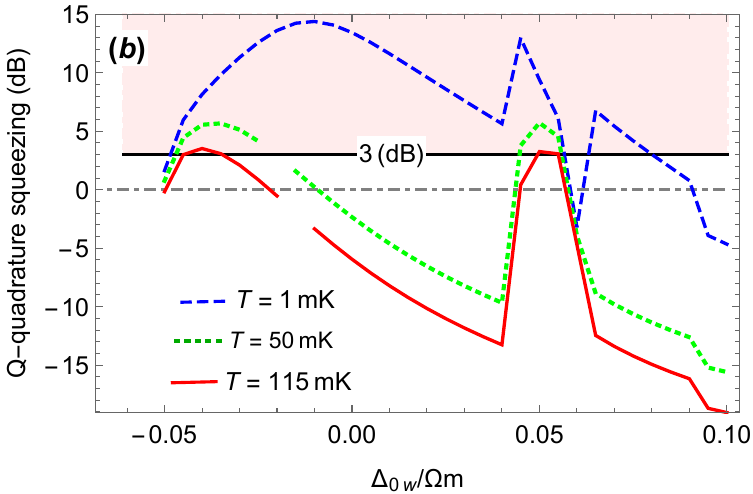}}
   \caption{(Color online) Squeezing degree of the $Q$-quadrature of the MO vs the dimensionless microwave detuning $\Delta_{0w}/\Omega_{m}$ for (a) different values of the QOC and (b) different temperatures. In panel (a) the green-dotted, blue-dashed, and red-solid lines, respectively, correspond to $g_{2}=0$, $g_{2}/g_{1}=10^{-3}$, and $g_{2}/g_{1}=4\times 10^{-3}$, and the temperature is $T=1$mK. In panel (b) the blue-dashed, green-dotted, and red-solid lines, respectively, correspond to $T=0.001$K, $T=0.05$K, and $T=0.115$K, and we have set $g_{2}/g_{1}=8\times 10^{-3}$. The shadowed pink region corresponds to squeezing beyond the 3 dB limit. Here we have set $\mathcal{P}=1 mW$ and $\mathcal{P}_{w}=1 mW$. The other parameters are the same as in Fig.~\ref{fig2}.}
   \label{fig12}
\end{figure} 

Thus the MO is said to be squeezed in the quadrature $Q$ ($P$) if $S_{Q}>0$ ($S_{P}>0$). Since 3dB squeezing, which corresponds to 50$\%$ noise reduction below the zero-point level (i.e., $\sigma_{j}=\sigma_{\rm zpf}/2$) is a limit for many proposals, the squeezing beyond 3dB can be regarded as strong squeezing. Mechanical squeezing beyond the 3dB limit has been experimentally realized by combining reservoir engineering and backaction-evading measurement in a microwave optomechanical system \cite{lei2016}.

Based on numerical calculations, we find that the $P$-quadrature fluctuations cannot be less than $1/2$, while the $Q$-quadrature fluctuations can be less than $1/2$. Therefore, in what follows we will focus our discussion only on the squeezing of the position quadrature. In Fig.~\ref{fig11a} we have plotted the squeezing degree of the $Q$-quadrature, $S_{Q}$, in dB as a function of $\Delta_{0w}/\Omega_{m}$ and $g_{2}/g_{1}$. As is seen, for the positive sign of QOC and in the vicinity of the microwave resonance frequency, $\Delta_{0w}/\Omega_{m}=0$, the 3-dB limit of squeezing can be beaten for the $Q$-quadrature of the MO so that it may be squeezed to high degrees (up to 15 dB). Nevertheless, mechanical squeezing is not possible for zero or negative values of QOC. In  Fig.~\ref{fig11b} we have shown the behavior of $S_{Q}$ versus the positive values of the ratio $g_{2}/g_{1}$ for $\Delta_{0w}/\Omega_{m}=0$. The figure shows that the mechanical squeezing ($S_{Q}>0$ dB) starts to appear at $g_{2}/g_{1}=0.001$, and strong mechanical squeezing($S_{Q} >$ 3dB)  is achievable for $0.0015<g_{2}/g_{1}<0.04$. We present the squeezing degree of the $Q$-quadrature with respect to $\Delta_{0w}/\Omega_{m}$ for three different non-negative values of $g_{2}/g_{1}$  in Fig.~\ref{fig12a}. As can be seen, in the absence of the QOC no squeezing is observed in the mechanical mode. However, for larger positive values of the QOC the beating of the standard squeezing limit of 3 dB is possible. Finally, we display in  Fig.~\ref{fig12b} the impact of the environment temperature $T$ on the squeezing degree of the $Q$ quadrature of the MO in the presence of QOC. Evidently, the degree of squeezing strongly decreases with increasing temperature. However, as is seen, the quadrature squeezing may survive up to $T=0.115$K for $g_{2}=8\times 10^{-3} g_{1}$.

\section{Conclusions}\label{sec6}
To conclude, we have investigated the effects of quadratic coupling between mechanical and optical modes on the steady-state bipartite entanglements, mechanical ground-state cooling, and mechanical quadrature squeezing in a hybrid electro-optomechanical system. To do this, we assumed that the MO is linearly coupled to the microwave mode, while it is simultaneously coupled to the optical mode through both LOC and QOC. Here, we summarize the main advantageous features of the model under consideration: (i) In the presence of QOC the steady-state entanglement between optical and mechanical modes can be amplified by about 2 order of magnitude around the temperature of 1mK. In addition, the QOC can make the enhanced optical-mechanical entanglement robust against the thermal noise, such that it survives up to 0.1K. (ii) In the optical and microwave red-detuned regime, the mechanical ground-state cooling can be enhanced by about 1 order of magnitude through tuning the QOC parameter. (iii) In the presence of QOC with a positive sign, the 3-dB limit of mechanical squeezing can be surpassed up to 15dB in the vicinity of the microwave resonance frequency. The generated mechanical squeezing may survive up to the temperature $T= 0.115$K .

\appendix

\section{Derivation of Hamiltonian (2)}
    \label{appendix1}
In this appendix we drive the Hamiltonian \eqref{eq2}. The original Hamiltonian of the considered electro-optomechanical system is 
\begin{eqnarray}
   \label{eq 19}
\hat{H}=&&\hbar \omega_{c} \hat{a}^{\dagger} \hat{a} +\dfrac{\hbar \Omega_{m}}{2}(\hat{Q}^{2}+\hat{P}^{2})+\dfrac{\hat{\phi}^{2}}{2L}+\dfrac{\hat{q}^{2}}{2[C+C_{0}(x)]}-e(t) \hat{Q}\nonumber \\
&&-\hbar g_{1}\hat{a}^{\dagger}\hat{a}\hat{Q}-\hbar g_{2}\hat{a}^{\dagger}\hat{a}\hat{Q}^{2}-i\hbar E_{d}(\hat{a}e^{i\omega_{d}t}-\hat{a}^{\dagger}e^{-i\omega_{d}t}),
\end{eqnarray}
where $(\hat{Q},\hat{P})$ are the canonical position and momentum of the MO with natural frequency $\Omega_{m}$, and $(\hat{\phi},\hat{q})$ are the canonical coordinates for the microwave cavity. $\hat{\phi}$ denotes the flux through an equivalent inductor $L$, and $\hat{q}$ describes the charge on an equivalent capacitor $C$. Furthermore, $(\hat{a},\hat{a}^{\dagger})$ are the creation an annihilation of the optical mode. The coherent driving of the  microwave cavity is given by the electric potential $e(t)=-i\sqrt{2\hbar \omega_{w}L}E_{dw}(e^{i\omega_{dw}t}-e^{-i\omega_{dw}t})$. The coupling of the MO with the microwave mode is due to the dependence of the capacitor capacity, $C_{0}(Q)$, on the displacement $Q$ of the mechanical membrane. We can expand $C_{0}(Q)$ around the equilibrium position of the LC circuit, corresponding to unperturbed distance $d$ between the parallel plates of the capacitor, with corresponding bare capacitance $C_{0}$, $C_{0}(Q)=C_{0}(1+Q(t)/d)$, where $Q(t)$ is the small displacement of the membrane around its equilibrium position. By using the Taylor expansion of the capacitive energy, we obtain to first order with respect to $Q(t)$

\begin{equation}
   \label{eq 20}
 \dfrac{q^{2}}{2[C+C_{0}(Q)]}=\dfrac{q^{2}}{2C_{\Sigma}}-\dfrac{\mu}{2dC_{\Sigma}}Q(t)q^{2},
\end{equation}
where $C_{\Sigma}=C+C_{0}$ and $\mu =\dfrac{C_{0}}{C_{\Sigma}}$. Now, we can rewrite the Hamiltonian of Eq.\eqref{eq 19} in terms of the creation and annihilation operators of the microwave mode $\hat{a}^{\dagger}_{w}$, and $\hat{a}_{w}$ ($[\hat{a}_{w},\hat{a}^{\dagger}_{w}]=1$), and the dimensionless position and momentum operators of the MO as
\begin{eqnarray}
   \label{eq 21}
   \hat{H}&&=\hbar \omega_{c}\hat{a}^{\dagger}\hat{a}+\dfrac{\hbar \Omega_{m}}{2}(\hat{Q}^{2}+\hat{P}^{2})+\hbar \omega_{w}\hat{a}^{\dagger}_{w}\hat{a}_{w}\nonumber \\
   &&-\hbar g_{1}\hat{a}^{\dagger}\hat{a}\hat{Q}-\hbar g_{2}\hat{a}^{\dagger}\hat{a}\hat{Q}^{2}-\dfrac{\hbar}{2}g_{w}(\hat{a}_{w}+\hat{a}^{\dagger}_{w})^{2}\hat{Q}\nonumber \\
   &&-i\hbar E_{dw}(e^{i\omega_{dw}t}-e^{-i\omega_{dw}t})(\hat{a}_{w}+\hat{a}^{\dagger}_{w})-i\hbar E_{d}(\hat{a}e^{i\omega_{d}t}-\hat{a}^{\dagger}e^{-i\omega_{d}t}),\nonumber \\   
\end{eqnarray}

where 
\begin{eqnarray}
 &&  \hat{a}_{w}=\sqrt{\dfrac{\omega_{w}L}{2\hbar}}\hat{q}+\dfrac{i}{\sqrt{2\hbar \omega_{w}L}}\hat{\phi},\nonumber \\
 && g_{w}=\dfrac{\mu \omega_{w}}{2d}x_{\rm zpf}.
\end{eqnarray}

Using the interaction picture with respect to $\hat{H}_{o}=\hbar  \omega_{d}\hat{a}^{\dagger}\hat{a}+\hbar \omega_{dw}\hat{a}^{\dagger}_{w}\hat{a}_{w}$ and neglecting the fast oscillating terms at $\pm2\omega_{d}$, and $\pm2\omega_{dw}$, the Hamiltonian of the system becomes
\begin{eqnarray}
&&\!\!\!\!\!\!\!\!\!\! \hat{H}=\hbar \Delta_{0c}\hat{a}^{\dagger} \hat{a} +\dfrac{\hbar \Omega_{m}}{2}(\hat{Q}^{2}+\hat{P}^{2})+\hbar \Delta_{0w} \hat{a}_{w}^{\dagger}\hat{a}_{w} \nonumber\\
&& - \hbar g_{1}\hat{a}^{\dagger} \hat{a} \hat{Q} -\hbar g_{2} \hat{a}^{\dagger} \hat{a} \hat{Q}^{2} - \hbar g_{w} \hat{a}_{w}^{\dagger} \hat{a}_{w} \hat{Q}\nonumber\\
&&-i\hbar E_{d}(\hat{a}-\hat{a}^{\dagger})-i\hbar\,E_{dw}(\hat{a}_{w}-\hat{a}_{w}^{\dagger}),
\end{eqnarray}
where $\Delta_{0c}=\omega_{c}-\omega_{d}$ and $\Delta_{0w}=\omega_{w}-\omega_{dw}$ are, respectively, the bare detunings of the optical and microwave modes.

\section{\textbf{ Linearized quantum Langevian equations}}
    \label{appendix2}
The linearized quantum Langevin equations for the fluctuation operators read as

\begin{subequations}
	\label{le}
	\begin{align}
	&\delta \dot{\hat{a}}=-(i\Delta_{c}+\kappa)\delta \hat{a}+i \tilde{G}\langle a \rangle_{ss} \delta \hat{Q}+\sqrt{2\kappa}\delta \hat{a}_{in},\\
	&\delta \dot{\hat{Q}}=\Omega_{m}\delta \hat{P}, \\
	&\delta \dot{\hat{P}}=-\tilde{\Omega}_{m} \delta \hat{Q}+\tilde{G}\langle a \rangle_{ss} (\delta \hat{a}+\delta \hat{a}^{\dagger})+g_{w} \langle a_{w} \rangle_{ss}(\delta \hat{a}_{w}+\delta \hat{a}_{w}^{\dagger})\nonumber \\
	&\qquad -\gamma_{m}\delta P+\sqrt{2\gamma_{m}}\delta P_{in} ,\\
	&\delta \dot{\hat{a}}_{w}=-(i\Delta_{w}+\kappa_{w})\delta \hat{a}_{w}+ig_{w}\langle a_{w} \rangle_{ss}\delta \hat{Q}+\sqrt{2\kappa_{w}}\delta \hat{a}_{in,w},
	\end{align}
\end{subequations}
where $\tilde{G}=g_{1}+2g_{2}\langle\,Q\rangle_{ss}$ is the effective single-photon optomechanical coupling strength, modified by the QOC. Here, without loss of generality, we can take $\langle a \rangle_{ss}$ and $\langle a_{w} \rangle_{ss}$ to be real by appropriate choice of the phases of the driving fields $E_{d}$ and $E_{dw}$. By introducing the dimensionless optical-field quadratures fluctuations as $\delta \hat{Q}_{o}=\dfrac{\delta \hat{a}+\delta \hat{a}^{\dagger}}{2}$ and $\delta \hat{P}_{o}=\dfrac{\delta \hat{a}-\delta \hat{a}^{\dagger}}{2 i}$, and also the dimensionless microwave-field quadratures fluctuations as $\delta \hat{Q}_{w}=\dfrac{\delta \hat{a}_{w}+\delta \hat{a}_{w}^{\dagger}}{2}$ and $\delta \hat{P}_{w}=\dfrac{\delta \hat{a}_{w}-\delta \hat{a}_{w}^{\dagger}}{2 i}$, the set of Eqs.~\eqref{le} can be expressed in the compact matrix form
\begin{equation}
\label{eq9}
\delta \dot{\hat u}(t)\,=\,\textbf{A}\,\delta \hat u(t)\,+\,\delta \hat n\,(t)\,,
\end{equation}
where
\begin{equation}
\label{u}
 \delta \hat u=(\delta \hat{Q}_{o} , \delta \hat{P}_{o},\delta\hat {Q} ,\delta \hat{P} , \delta \hat{Q}_{w} ,\delta \hat{P}_{w})^{T}
\end{equation}
is the vector of continuous-variable fluctuations and $\textbf{A}$ is the $6\times 6$ drift matrix which is given by
\begin{equation}
  \label{A}
 \textbf{A} =  
 \begin{pmatrix}
 {-\kappa} & {\Delta_{c}} & {0} & {0} & {0} & {0} \\
{-\Delta_{c}} & {-\kappa} & {\tilde{G}\langle Q_{o} \rangle_{ss}} & {0} & {0} & {0} \\
{0} & {0} & {0} & {\Omega_{m}} & {0} & {0} \\
{\tilde{G}\langle Q_{o} \rangle_{ss}} & {0} & {-\tilde{\Omega}_{m}} & {-\gamma_{m}} & {g_{w} \langle Q_{w} \rangle_{ss}} & {0} \\
{0} & {0} & {0} & {0} & {-\kappa_{w}} & {\Delta_{w}} \\
{0} & {0} & {g_{w}\langle Q_{w} \rangle_{ss}} & {0} & {-\Delta_{w}} & {-\kappa_{w}} 
\end{pmatrix} ,
\end{equation}
and
\begin{eqnarray} 
 \label{n}
 \delta \hat{n}(t)=&(\sqrt{2\kappa}\delta\hat{Q}_{in,o} ,\sqrt{2\kappa}\delta\hat{P}_{in,o} ,0 , \sqrt{2\gamma_{m}}\delta \hat{P}_{in},\sqrt{2\kappa_{w}}\delta \hat{Q}_{in,w}\nonumber \\
& ,\sqrt{2\kappa_{w}}\delta\hat{P}_{in,w})^{T}
\end{eqnarray}
represents the corresponding vector of noises.

\section{Stability conditions}
   \label{appendix3}
 
The explicit expressions of the stability conditions for the system under study, which are obtained by applying the Routh–Hurwitz criterion, are as follows
\begin{widetext}
\small
\begin{subequations}

    \begin{eqnarray}
&s_{1}\equiv \gamma_{m}+2\kappa +2\kappa_{w}>0, \\
\quad \nonumber\\
&s_{2}\equiv 2\gamma_{m}^{2}(\kappa +\kappa_{w})+4\gamma_{m}(\kappa +\kappa_{w})^{2}+2(\Delta_{c}^{2}\kappa +\Delta_{w}^{2}\kappa_{w}+(\kappa +\kappa_{w})(\kappa^{2}+3\kappa \kappa_{w}+\kappa_{w}^{2}))+\gamma_{m}\Omega_{m}\tilde{\Omega}_{m}>0,\\
\quad \nonumber \\
&s_{3}\equiv 4\kappa \kappa_{w}((\Delta_{c}-\Delta_{w})^{2}+(\kappa +\kappa_{w})^{2})((\Delta_{c}+\Delta_{w})^{2}+(\kappa +\kappa_{w})^{2})+2\gamma^{3}(\Delta_{c}^{2}\kappa +\Delta_{w}^{2}\kappa_{w}+(\kappa +\kappa_{w})(\kappa^{2}+3\kappa \kappa_{w}+\kappa_{w}^{2}))+4(\kappa +\kappa_{w})(G^{2}\Delta_{c}\kappa +G_{w}^{2}\Delta_{w}\kappa_{w})\Omega_{m}\nonumber\\  &+\gamma_{m}^{2}(4\kappa^{4}+4\Delta_{w}^{2}\kappa \kappa_{w}+20\kappa^{3}\kappa_{w}+4\Delta_{w}^{2}\kappa_{w}^2+32\kappa^{2}\kappa_{w}^{2}+20\kappa \kappa_{w}^{3}+4\kappa_{w}^{4}+4\Delta_{c}^{2}\kappa (\kappa +\kappa_{w})+G^{2}\Delta_{c}\Omega_{m}+G_{w}^{2}\Delta_{w}\Omega_{m}+4(\kappa +\kappa_{w})^{2}\Omega_{m}\tilde{\Omega}_{m})+2\gamma_{m}(\Delta_{c}^{4}\kappa \nonumber \\
&+\kappa^{5}+\Delta_{w}^{4}\kappa_{w}+4\Delta_{w}^{2}\kappa^{2}\kappa_{w}+8\kappa^{4}\kappa_{w}+8\Delta_{w}^{2}\kappa \kappa_{w}^{2}+20\kappa^{3}\kappa_{w}^{2}+2 \Delta_{w}^{2}\kappa_{w}^{3}+20\kappa^{2}\kappa_{w}^{3}+8\kappa\kappa_{w}^{4}+\kappa_{w}^{5}+G_{w}^{2}\Delta_{w}\kappa \Omega_{m}+2G_{w}^{2}\Delta_{w}\kappa_{w}\Omega_{m}+G^{2}\Delta_{c}(2\kappa +\kappa_{w})\Omega_{m}\nonumber \\
&+2(-\Delta_{w}^{2}\kappa_{w}+(\kappa +\kappa_{w})(\kappa^{2}+\kappa \kappa_{w}+\kappa_{w}^{2}))\Omega_{m}\tilde{\Omega}_{m}+(\kappa +\kappa_{w})\Omega_{m}^{2}\tilde{\Omega}_{m}^{2}+2\Delta_{c}^{2}\kappa(\kappa^{2}+4\kappa \kappa_{w}+2\kappa_{w}^{2}-\Omega_{m}\tilde{\Omega}_{m}))>0,\\
\quad  \nonumber \\
&s_{4}\equiv (\gamma_{m} (\Delta_{c}^{2} +\kappa^{2}) (\Delta_{w}^{2} +\kappa_{w}^2) -2 (G_{w}^{2}\Delta_{w} \kappa +G^{2} \Delta_{c}\kappa_{w})\Omega_{m} +2 (\Delta_{w}^{2}\kappa +\kappa_{w}(\Delta_{c}^{2}+\kappa(\kappa +\kappa_{w}))\Omega_{m}\tilde{\Omega}_{m})(-\gamma_{m} \Delta_{c}^{2}\Delta{w}^{2}-\gamma_{m}\Delta_{w}^{2}\kappa^{2} -\gamma_{m}\Delta_{c}^{2}\kappa_{w}^{2} -\gamma_{m}\kappa^{2}\kappa_{w}^{2}\nonumber \\
&+2G_{w}^{2}\Delta_{w}\kappa \Omega_{m} +2 G^{2}\Delta_{c}\kappa_{w}\Omega_{m}-2\Delta_{w}^{2}\kappa \Omega_{m}\tilde{\Omega}_{m}-2\Delta_{c}^{2}\kappa_{w}\Omega_{m}\tilde{\Omega}_{m}-2\kappa^{2}\kappa_{w}\Omega_{m} \tilde{\Omega}_{m}-2\kappa\kappa_{w}^{2}\Omega_{m}\tilde{\Omega}_{m}+(\Delta_{c}^{2}+\Delta_{w}^{2}+\kappa^{2}+4\kappa \kappa_{w}+\kappa_{w}^{2} +2\gamma_{m} (\kappa +\kappa_{w}) +\Omega_{m}\tilde{\Omega}_{m})\nonumber \\
& (\gamma_{m}(\Delta_{c}^{2}+\Delta_{w}^{2}+\kappa^{2}+4\kappa \kappa_{w}+\kappa_{w}^{2})+2(\Delta_{w}^{2}\kappa +\Delta_{c}^{2}\kappa_{w}+(\kappa +\kappa_{w})(\kappa \kappa_{w}+\Omega_{m}\tilde{\Omega}_{m}))))-(\Delta_{w}^{2} \kappa^{2}+\kappa^{2}\kappa_{w}^{2}+2\gamma_{m}\kappa (\Delta_{w}^{2}+\kappa_{w}(\kappa +\kappa_{w}))-G^{2}\Delta_{c} \Omega_{m}\nonumber \\
&-G_{w}^{2}\Delta_{w}\Omega_{m}+(\Delta_{w}^{2}+\kappa^{2}+4\kappa \kappa_{w}+\kappa_{w}^{2})\Omega_{m}\tilde{\Omega}_{m}+\Delta_{c}^{2}(\Delta_{w}^{2}+2 \gamma_{m}\kappa_{w}+\kappa_{w}^{2}+\Omega_{m}\tilde{\Omega}_{m})) (-(\gamma_{m}+2 (\kappa +\kappa_{w})) (\gamma_{m} (\Delta_{c}^{2} + \kappa^{2}) (\Delta_{w}^{2}+\kappa_{w}^{2})-2(G_{w}^{2} \Delta_{w}\kappa \nonumber \\
&+G^{2}\Delta_{c}\kappa_{w}) \Omega_{m}+2(\Delta_{w}^{2}\kappa +\kappa_{w}(\Delta_{c}^{2}+\kappa(\kappa +\kappa_{w})))\Omega_{m}\tilde{\Omega}_{m})+(\gamma_{m} (\Delta_{c}^{2}+\Delta_{w}^{2}+\kappa^{2}+4\kappa \kappa_{w}+\kappa_{w}^{2})+2(\Delta_{w}^{2}\kappa +\Delta_{c}^{2}\kappa_{w}+(\kappa +\kappa_{w}) (\kappa \kappa_{w}+\Omega_{m}\tilde{\Omega}_{m})))^{2})\nonumber \\
& +(\gamma_{m}+2 (\kappa +\kappa_{w})) (\Omega_{m} (2 \gamma_{m}^{2}(\kappa +\kappa_{w})+4\gamma_{m}(\kappa +\kappa_{w})^{2}+2(\Delta_{c}^{2}\kappa +\Delta_{w}^{2}\kappa_{w}+(\kappa +\kappa_{w}) (\kappa^{2}+3\kappa \kappa_{w}+\kappa _{w}^{2}))+\gamma_{m}\Omega_{m}\tilde{\Omega}_{m})(-G_{w}^{2}\Delta_{w}(\Delta_{c}^{2}+\kappa^{2})\nonumber \\
&-(\Delta_{w}^{2}+\kappa_{w}^{2})(G^{2}\Delta_{c}-(\Delta_{c}^{2}+\kappa^{2})\tilde{\Omega}_{m}))-(\gamma_{m}(\Delta_{c}^{2}+\kappa^{2})(\Delta_{w}^{2}+\kappa_{w}^{2})-2(G_{w}^{2}\Delta_{w}\kappa +G^{2}\Delta_{c}\kappa_{w})\Omega_{m}+2(\Delta_{w}^{2}\kappa +\kappa_{w}(\Delta_{c}^{2}+\kappa (\kappa +\kappa_{w}))\Omega_{m}\tilde{\Omega}_{m})\nonumber \\
&(-\Delta_{c}^{2}\Delta_{w}^{2}-2\gamma_{m}\Delta_{w}^{2}\kappa -\Delta_{w}^{2}\kappa^{2}-2\gamma_{m}\Delta_{c}^{2}\kappa_{w}-2\gamma_{m}\kappa^{2}\kappa_{w}-\Delta_{c}^{2}\kappa_{w}^{2}-2\gamma_{m}\kappa \kappa_{w}^{2}-\kappa^{2}\kappa_{w}^{2}+G^{2}\Delta_{c}\Omega_{m}+G_{w}^{2}\Delta_{w}\Omega_{m}-\Delta_{c}^{2}\Omega_{m}\tilde{\Omega}_{m} -\Delta_{w}^{2}\Omega_{m}\tilde{\Omega}_{m}-\kappa^{2}\Omega_{m}\nonumber \\
& \tilde{\Omega}_{m}-4\kappa \kappa_{w}\Omega_{m}\tilde{\Omega}_{m}-\kappa_{w}^{2}\Omega_{m}\tilde{\Omega}_{m}+(\Delta_{c}^{2}+\Delta_{w}^{2}+\kappa^{2}+4\kappa \kappa_{w}+\kappa_{w}^{2}+2\gamma_{m}(\kappa +\kappa_{w})+\Omega_{m}\tilde{\Omega}_{m})^{2}) +(\Delta_{w}^{2}\kappa^{2}+\kappa^{2}\kappa_{w}^{2}+2\gamma_{m}\kappa (\Delta_{w}^{2}+\kappa_{w}(\kappa +\kappa_{w}))\nonumber \\
&-G^{2}\Delta_{c}\Omega_{m}-G_{w}^{2}\Delta_{w}\Omega_{m}+(\Delta_{w}^{2}+\kappa^{2}+4\kappa \kappa_{w}+\kappa_{w}^{2})\Omega_{m} \tilde{\Omega}_{m}+\Delta_{c}^{2}(\Delta_{w}^{2}+2\gamma_{m}\kappa_{w}+\kappa_{w}^{2}+\Omega_{m}\tilde{\Omega}_{m}))(-(\gamma_{m} +2(\kappa +\kappa_{w}))(\Delta_{w}^{2}\kappa^{2}+\kappa^{2}\kappa_{w}^{2}+2\gamma_{m}\kappa(\Delta_{w}^{2}+\kappa_{w}\nonumber \\
&(\kappa +\kappa_{w}))-G^{2}\Delta_{c}\Omega_{m}-G_{w}^{2}\Delta_{w}\Omega_{m}+(\Delta_{w}^{2} +\kappa^{2}+4\kappa \kappa_{w}+\kappa_{w}^{2})\Omega_{m}\tilde{\Omega}_{m}+\Delta_{c}^{2}(\Delta_{w}^{2}+2\gamma_{m}\kappa_{w}+\kappa_{w}^{2}+\Omega_{m}\tilde{\Omega}_{m}))+(\Delta_{c}^{2}+\Delta_{w}^{2}+\kappa^{2}+4\kappa \kappa_{w}+\kappa_{w}^{2}\nonumber \\
&+2\gamma_{m}(\kappa +\kappa_{w})+\Omega_{m}\tilde{\Omega}_{m})(\gamma_{m}(\Delta_{c}^{2}+\Delta_{w}^{2}+\kappa^{2}+4\kappa \kappa_{w}+\kappa_{w}^{2})+2(\Delta_{w}^{2}\kappa +\Delta_{c}^{2}\kappa_{w}+(\kappa +\kappa_{w})(\kappa \kappa_{w}+\Omega_{m}\Omega_{m})))))>0, \\
  \quad \nonumber \\
&s_{5}\equiv \Omega_{m}(G_{w}^{2}\Delta_{w}(\Delta_{c}^{2}+\kappa^{2})+(\Delta_{w}^{2}+\kappa_{w}^{2})(G^{2}\Delta_{c}-(\Delta_{c}^{2}+\kappa^{2})\tilde{\Omega}_{m}))(-(\gamma_{m}+2(\kappa +\kappa_{w}))((2\gamma_{m}^{2}(\kappa +\kappa_{w})+4\gamma_{m}(\kappa +\kappa_{w})^{2}+2 (\Delta_{c}^{2} \kappa + \Delta_{w}^{2}\kappa_{w} +(\kappa +\kappa_{w})\nonumber \\
&(\kappa^{2}+3\kappa \kappa_{w}+\kappa_{w}^{2}))+\gamma_{m}\Omega_{m}\tilde{\Omega}_{m})(\gamma_{m}(\Delta_{c}^{2}+\kappa^{2})(\Delta_{w}^{2}+\kappa_{w}^{2})-2 (G_{w}^{2}\Delta_{w}\kappa +G^{2}\Delta_{c}\kappa_{w}) \Omega_{m}+2(\Delta_{w}^{2}\kappa +\kappa_{w}(\Delta_{c}^{2}+\kappa (\kappa + \kappa_{w})))\Omega_{m}\tilde{\Omega}_{m})\nonumber \\
&-(\gamma_{m}+2(\kappa +\kappa_{w}))^{2}\Omega_{m}(-G_{w}^{2}\Delta_{w}(\Delta_{c}^{2}+\kappa^{2})-(\Delta_{w}^{2}+\kappa_{w}^{2})(G^{2}\Delta_{c}-(\Delta_{c}^{2}+\kappa^{2})\tilde{\Omega}_{m})))+(4\kappa \kappa_{w}((\Delta_{c}-\Delta_{w})^{2}+(\kappa +\kappa_{w})^{2})((\Delta_{c}+\Delta_{w})^{2}+(\kappa +\kappa_{w})^{2})\nonumber \\
&+2\gamma_{m}^{3}(\Delta_{c}^{2} \kappa +\Delta_{w}^{2}\kappa_{w}+(\kappa +\kappa_{w})(\kappa^{2}+3\kappa \kappa_{w}+ \kappa_{w}^{2}))+4 (\kappa +\kappa_{w})(G^{2}\Delta_{c}\kappa +G_{w}^{2}\Delta_{w} \kappa_{w})\Omega_{m}+\gamma_{m}^{2} (4\kappa^{4} + 4\Delta_{w}^{2}\kappa \kappa_{w}+20\kappa^{3}\kappa_{w}+4\Delta_{w}^{2}\kappa_{w}^{2}+32 \kappa^{2}\kappa_{w}^{2}\nonumber \\
&+20\kappa \kappa_{w}^{3}+4 \kappa_{w}^{4}+4\Delta_{c}^{2}\kappa (\kappa +\kappa_{w})+G^{2}\Delta_{c}\Omega_{m}+G_{w}^{2}\Delta_{w}\Omega_{m}+4(\kappa +\kappa_{w})^{2}\Omega_{m}\tilde{\Omega}_{m})+2\gamma_{m}(\Delta_{c}^{4}\kappa +\kappa^{5}+\Delta_{w}^{4}\kappa_{w}+4\Delta_{w}^{2}\kappa^{2}\kappa_{w}+8\kappa^{4 }\kappa_{w}+8 \Delta_{w}^{2} \kappa \kappa_{w}^{2}\nonumber \\
&+20\kappa^{3}\kappa_{w}^{2}+2\Delta_{w}^{2} \kappa_{w}^{3}+20 \kappa^{2}\kappa_{w}^{3}+8\kappa \kappa_{w}^{4}+\kappa_{w}^{5}+G_{w}^{2}\Delta_{w}\kappa \Omega_{m}+2 G_{w}^{2} \Delta_{w} \kappa_{w} \Omega_{m}+G^{2} \Delta_{c} (2 \kappa +\kappa_{w})\Omega_{m} +2(-\Delta_{w}^{2} \kappa_{w} +(\kappa + \kappa_{w})(\kappa^{2} +\kappa \kappa_{w} +\kappa_{w}^{2})) \Omega_{m}\tilde{\Omega}_{m}\nonumber \\
&+(\kappa + \kappa_{w})\Omega_{m}^{2} \tilde{\Omega}_{m}^{2}+2\Delta_{c}^{2} \kappa (\kappa^{2}+4\kappa \kappa_{w}+2\kappa_{w}^{2} -\Omega_{m}\tilde{\Omega}_{m})))(\gamma_{m}(\Delta_{c}^{2} +\Delta_{w}^{2}+\kappa^{2}+4\kappa \kappa_{w}+\kappa_{w}^{2})+2(\Delta_{w}^{2} \kappa +\Delta_{c}^{2} \kappa_{w}+(\kappa +\kappa_{w})(\kappa \kappa_{w}+\Omega_{m}\tilde{\Omega}_{m}))))\nonumber \\
&+(\gamma_{m} (\Delta_{c}^{2} +\kappa^{2})(\Delta_{w}^{2}+\kappa_{w}^{2})-2(G_{w}^{2}\Delta_{w}\kappa +G^{2}\Delta_{c} \kappa_{w}) \Omega_{m}+2 (\Delta_{w}^{2}\kappa +\kappa_{w} (\Delta_{c}^{2} +\kappa (\kappa + \kappa_{w}))) \Omega_{m}\tilde{\Omega}_{m})((\gamma_{m} (\Delta_{c}^{2} +\kappa^{2})(\Delta_{w}^{2}+\kappa_{w}^{2})-2(G_{w}^{2}\Delta_{w}\kappa \nonumber \\
& +G^{2}\Delta_{c}\kappa_{w})\Omega_{m}+2(\Delta_{w}^{2}\kappa +\kappa_{w}(\Delta_{c}^{2}+\kappa (\kappa + \kappa_{w}))) \Omega_{m}\tilde{\Omega}_{m})(-\gamma_{m}\Delta_{c}^{2} \Delta_{w}^{2}- \gamma_{m}\Delta_{w}^{2}\kappa^{2}- \gamma_{m}\Delta_{c}^{2}\kappa_{w}^{2}-\gamma_{m}\kappa^{2}\kappa_{w}^{2}+2 G_{w}^{2}\Delta_{w}\kappa \Omega_{m}+2 G^{2}\Delta_{c}\kappa_{w}\Omega_{m}\nonumber \\
&-2\Delta_{w}^{2}\kappa \Omega_{m}\tilde{\Omega}_{m}-2 \Delta_{c}^{2} \kappa_{w}\Omega_{m}\tilde{\Omega}_{m}-2 \kappa^{2}\kappa_{w}\Omega_{m}\tilde{\Omega}_{m}-2\kappa \kappa_{w}^{2}\Omega_{m}\tilde{\Omega}_{m}+(\Delta_{c}^{2} +\Delta_{w}^{2} +\kappa^{2}+4 \kappa \kappa_{w}+\kappa_{w}^{2}+2\gamma_{m}(\kappa + \kappa_{w})+\Omega_{m} \tilde{\Omega}_{m})\nonumber \\
&(\gamma_{m} (\Delta_{c}^{2}+\Delta_{w}^{2}+\kappa^{2}+4 \kappa \kappa_{w}+\kappa_{w}^{2})+2(\Delta_{w}^{2}\kappa + \Delta_{c}^{2}\kappa_{w}+(\kappa + \kappa_{w})(\kappa \kappa_{w}+\Omega_{m} \tilde{\Omega}_{m}))))-(\Delta_{w}^{2}\kappa^{2} +\kappa^{2} \kappa_{w}^{2}+2 \gamma_{m} \kappa (\Delta_{w}^{2}+\kappa_{w}(\kappa + \kappa_{w}))\nonumber \\
&-G^{2}\Delta_{c}\Omega_{m} -G_{w}^{2} \Delta_{w}\Omega_{m} +(\Delta_{w}^{2}+\kappa^{2} +4 \kappa \kappa_{w}+\kappa_{w}^{2})\Omega_{m}\tilde{\Omega}_{m}+\Delta_{c}^{2}(\Delta_{w}^{2}+2\gamma_{m}\kappa_{w}+\kappa_{w}^{2}+\Omega_{m}\tilde{\Omega}_{m}))(-(\gamma_{m}+2 (\kappa +\kappa_{w})) (\gamma_{m}(\Delta_{c}^{2}+\kappa^{2})(\Delta_{w}^{2}+\kappa_{w}^{2})\nonumber \\
&-2 (G_{w}^{2}\Delta_{w}\kappa+G^{2}\Delta_{c}\kappa_{w}) \Omega_{m}+2(\Delta_{w}^{2}\kappa +\kappa_{w}(\Delta_{c}^{2} +\kappa (\kappa +\kappa_{w})))\Omega_{m} \tilde{\Omega}_{m})+(\gamma_{m}(\Delta_{c}^{2}+\Delta_{w}^{2}+\kappa^{2}+4\kappa \kappa_{w}+\kappa_{w}^{2})+2(\Delta_{w}^{2} \kappa +\Delta_{c}^{2}\kappa_{w}\nonumber \\
&+(\kappa +\kappa_{w})(\kappa \kappa_{w}+\Omega_{m} \tilde{\Omega}_{m})))^2)+(\gamma_{m}+2(\kappa +\kappa_{w}))(\Omega_{m}(2\gamma_{m}^{2} (\kappa +\kappa_{w})+4\gamma_{m}(\kappa +\kappa_{w})^{2}+2(\Delta_{c}^{2} \kappa + \Delta_{w}^{2}\kappa_{w}+(\kappa +\kappa_{w})(\kappa^{2}+3\kappa \kappa_{w}+\kappa_{w}^{2}))\nonumber \\
&+\gamma_{m}\Omega_{m}\tilde{\Omega}_{m})(-G_{w}^{2}\Delta_{w}(\Delta_{c}^{2}+\kappa^{2})-(\Delta_{w}^{2} +\kappa_{w}^{2})(G^{2} \Delta_{c}-(\Delta_{c}^{2}+\kappa^{2})\tilde{\Omega}_{m}))-(\gamma_{m}(\Delta_{c}^{2}+\kappa^{2})(\Delta_{w}^{2}+\kappa_{w}^{2})-2 (G_{w}^{2}\Delta_{w}\kappa + G^{2} \Delta_{c}\kappa_{w})\Omega_{m}\nonumber \\
&+2 (\Delta_{w}^{2} \kappa +\kappa_{w}(\Delta_{c}^{2}+\kappa (\kappa +\kappa_{w})))\Omega_{m}\tilde{\Omega}_{m})(-\Delta_{c}^{2}\Delta_{w}^{2}-2\gamma_{m}\Delta_{w}^{2}\kappa -\Delta_{w}^{2} \kappa^{2}-2\gamma_{m}\Delta_{c}^{2}\kappa_{w} -2\gamma_{m}\kappa^{2} \kappa_{w}-\Delta_{c}^{2}\kappa_{w}^{2}-2 \gamma_{m} \kappa \kappa_{w}^{2}-\kappa^{2}\kappa_{w}^{2}+ G^{2}\Delta_{c}\Omega_{m}\nonumber \\
&+G_{w}^{2}\Delta_{w}\Omega_{m}-\Delta_{c}^{2}\Omega_{m}\tilde{\Omega}_{m} -\Delta_{w}^{2}\Omega_{m}\tilde{\Omega}_{m}-\kappa^{2} \Omega_{m}\tilde{\Omega}_{m}-4 \kappa \kappa_{w}\Omega_{m}\tilde{\Omega}_{m}-\kappa_{w}^{2}\Omega_{m}\tilde{\Omega}_{m}+(\Delta_{c}^{2}+\Delta_{w}^{2}+\kappa^{2}+4\kappa \kappa_{w}+\kappa_{w}^{2}+2\gamma_{m}(\kappa +\kappa_{w})+\Omega_{m} \tilde{\Omega}_{m})^{2})\nonumber \\
&+(\Delta_{w}^{2}\kappa^{2}+\kappa^{2}\kappa_{w}^{2}+2\gamma_{m}\kappa (\Delta_{w}^{2}+\kappa_{w}(\kappa +\kappa_{w})) - G^{2}\Delta_{c}\Omega_{m}- G_{w}^{2}\Delta_{w}\Omega_{m}+(\Delta_{w}^{2}+\kappa^{2}+4 \kappa \kappa_{w}+\kappa_{w}^{2})\Omega_{m}\tilde{\Omega}_{m}+\Delta_{c}^{2}(\Delta_{w}^{2}+2\gamma_{m}\kappa_{w}+\kappa_{w}^{2}+\Omega_{m}\tilde{\Omega}_{m}))\nonumber \\
&(-(\gamma_{m} +2(\kappa +\kappa_{w}))(\Delta_{w}^{2}\kappa^{2}+\kappa^{2}\kappa_{w}^{2}+2 \gamma_{m}\kappa (\Delta_{w}^{2}+\kappa_{w}(\kappa +\kappa_{w}))- G^{2}\Delta_{c}\Omega_{m}-G_{w}^{2}\Delta_{w}\Omega_{m}+(\Delta_{w}^{2} +\kappa^{2}+4\kappa \kappa_{w}+\kappa_{w}^{2})\Omega_{m}\tilde{\Omega}_{m}+\Delta_{c}^{2} (\Delta_{w}^{2}\nonumber \\
&+2 \gamma_{m}\kappa_{w}+\kappa_{w}^{2}+\Omega_{m}\tilde{\Omega}_{m}))+(\Delta_{c}^{2}+\Delta_{w}^{2} +\kappa^{2}+4\kappa \kappa_{w}+\kappa_{w}^{2}+2\gamma_{m}(\kappa + \kappa_{w}) +\Omega_{m}\tilde{\Omega}_{m}) (\gamma_{m}(\Delta_{c}^{2}+\Delta_{w}^{2} +\kappa^{2}+4\kappa \kappa_{w}+\kappa_{w}^{2})+2 (\Delta_{w}^{2}\kappa +\Delta_{c}^{2}\kappa_{w}\nonumber \\
&+ (\kappa +\kappa_{w})(\kappa \kappa_{w}+\Omega_{m} \tilde{\Omega}_{m}))))))>0,\\
\quad \nonumber \\
&s_{6}\equiv -G_{w}^{2}\Delta_{c}^{2}\Delta_{w} \Omega_{m}-G^{2}\Delta_{c}\Delta_{w}^{2}\Omega_{m}-G_{w}^{2}\Delta_{w}\kappa^{2}\Omega_{m}- G^{2} \Delta_{c}\kappa_{w}^{2}\Omega_{m}+\Delta_{c}^{2}\Delta_{w}^{2} \Omega_{m}\tilde{\Omega}_{m}+\Delta_{w}^{2}\kappa^{2}\Omega_{m}\tilde{\Omega}_{m}+\Delta_{c}^{2} \kappa_{w}^{2} \Omega_{m}\tilde{\Omega}_{m}+\kappa^{2}\kappa_{w}^{2}\Omega_{m}\tilde{\Omega}_{m}>0.
   \end{eqnarray}
\end{subequations}
 \end{widetext}
 
Throughout this work, we have checked numerically that the chosen parameters satisfy these six independent stability conditions.

 \section{Ground-state cooling and energy equipartition}
   \label{appendix4}
      \begin{figure*}
	\includegraphics[width=17.5cm]{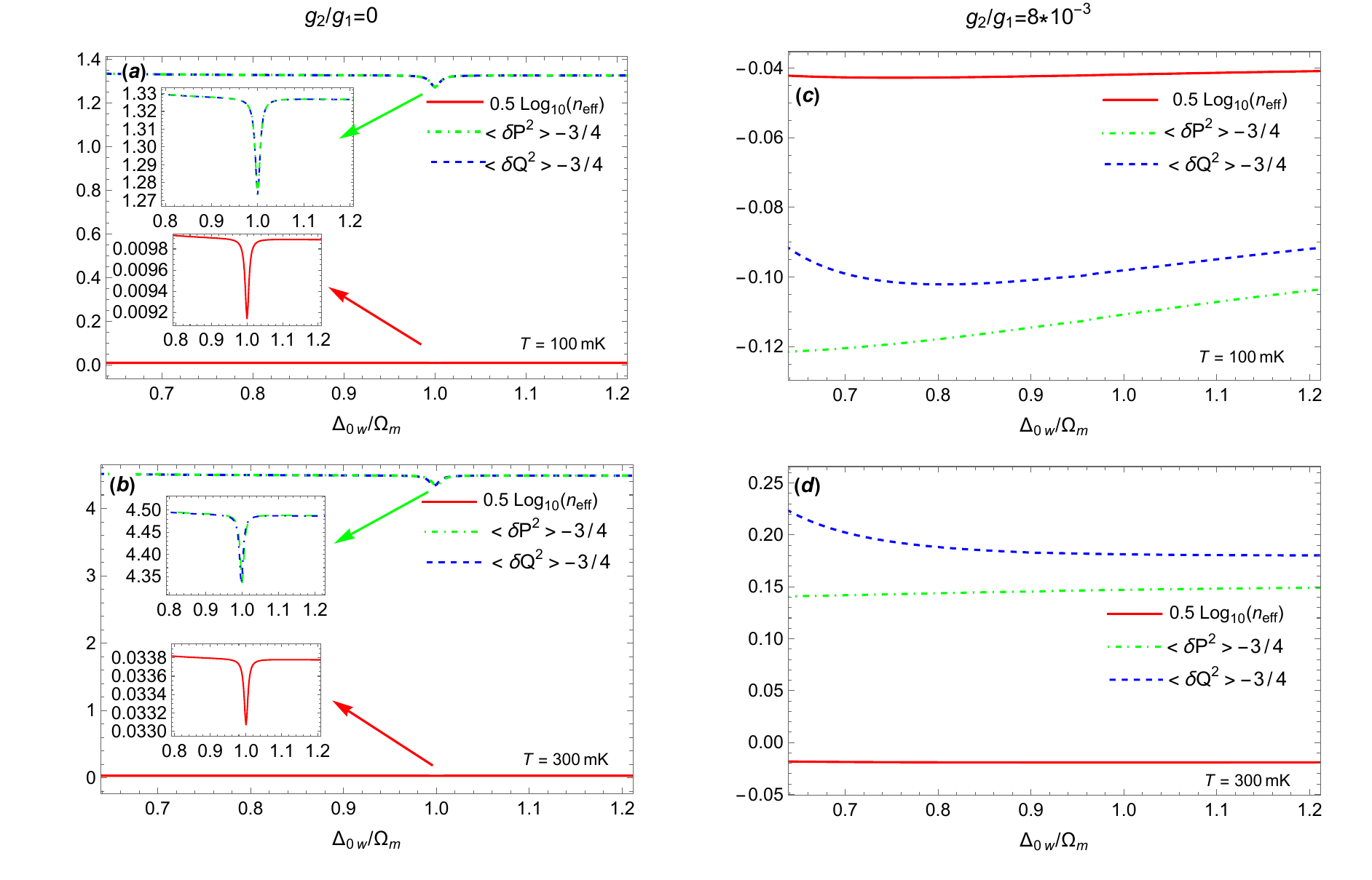}
	\caption{(Color online) Plots of effective mean phonon number $n_{\rm eff}$ (on logarithmic scale) and the variances $\langle \delta \hat{Q}^{2}\rangle-3/4$ and $\langle\hat{P}^{2}\rangle-3/4$ versus the normalized detuning $\Delta_{0w}/\Omega_{m}$ for $g_{2}=0$ (panels a and b) and $g_{2}=8\times10^{-3}g_{1}$ (panels c and d), and for different values of the environment temperature $T$. Other parameters are the same as in Fig.~\ref{fig2}.}
	\label{fig13}
\end{figure*}
    We have examined the condition of energy  equipartition $\langle \delta \hat{P}^{2}\rangle \approx \langle \delta \hat{Q}^{2}\rangle \approx 1/2$ in Fig.~\ref{fig13}, where we have plotted the effective mean phonon number $n_{\rm eff}$ (on logarithmic scale) and the variances $\langle \delta \hat{Q}^{2} \rangle -3/4$ and $\langle \delta \hat{P}^{2} \rangle-3/4$ versus $\Delta_{0w}/\Omega_{m}$  for $g_{2}=0$ (panels a and b) and $g_{2}=8\times10^{-3}g_{1}$ (panels c and d), and for different values of temperature $T=0.1 \rm K$ and $0.3 \rm K$. As can be seen, in the absence of QOC, energy equipartition breaks down, i.e., one has $\langle \delta \hat{P}^{2}\rangle \approx \langle \delta \hat{Q}^{2}\rangle > 1/2$ so that $n_{\rm eff}>1$ [see Figs.~\ref{fig13}(a) and \ref{fig13}(b)]. On the other hand, in the presence of QOC we have both $n_{\rm eff}<1$ and $\langle \delta \hat{P}^{2}\rangle \approx \langle \delta \hat{Q}^{2}\rangle \approx 1/2$ until the temperature $T=0.1 \rm K$. According to Fig.~\ref{fig13}(d) we understand that as the temperature $T$ increases up to $0.3 \rm K$  the ground-state cooling requirement $n_{\rm eff}<1$ is still satisfied, though the energy equipartition breaks down.


\begin{thebibliography}{79}%
  \setlength{\baselineskip}{0.2 \baselineskip}
  
\bibitem{aspelmeyer2014}M. Aspelmeyer, T. J. Kippenberg, and F. Marquardt, ``Cavity optomechanics," 
   \href{https://doi.org/10.1103/RevModPhys.86.1391}{Rev.Mod.Phys. \textbf{86}, 1391 (2014)}.
	
\bibitem{bowen2015} W. Bowen and G. Milburn, \textit{Quantum Optomechanics}
   \href{https://doi.org/10.1201/b19379}{(CRC Press, 2015)}.

\bibitem{teufel2011}J.D. Teufel, T. Donner, D. Li, J.W. Harlow, MS. Allman, K. Cicak, A.J. Sirois, J.D. Whittaker, K.W. Lehnert, and R.W. Simmonds, ``Sideband cooling of micromechanical motion to the quantum ground state,"
   \href{https://doi.org/10.1038/nature10261}{Nature \textbf{475}, 359 (2011)}.

\bibitem{connell2010} A. D. O' Connell, M. Hofheinz, M. Ansmann, R. C. Bialczak, M. Lenander, E. Lucero, M. Neeley, D. Sank, H. Wang, M. Weides, J. Wenner, J. M. Martinis, and A. N. Cleland, ``Quantum ground state and single-phonon control of a mechanical resonator," 
   \href{https://doi.org/10.1038/nature08967}{Nature. \textbf{464}, 697 (2010)}.
	
\bibitem{chan2011} J. Chan, T. P. M. Alegre, A. H. Safavi-Naeini, J. T. Hill, A. Krause, S. Groblacher, M. Aspelmeyer, and O. Painter, ``Laser cooling of a nanomechanical oscillator into its quantum ground state," 
   \href{https://doi.org/10.1038/nature10461}{Nature. \textbf{478}, 89 (2011)}.
	
\bibitem{hammerer2014}K. Hammerer, C. Genes, D. Vitali, P. Tombesi, G. Milburn, CH. Simon, and D. Bouwmeester, ``\textit{Nonclassical States of Light and Mechanics}"
   \href{https://doi.org/10.1007/978-3-642-55312-7_3}{(Springer, New York, 2014)}.
	

\bibitem{pontin2016} A. Pontin, M. Bonaldi, A. Borrielli, L. Marconi, F. Marino, G. Pandraud, G. A. Prodi,
	P. M. Sarro, E. Serra, and F. Marin, ``Dynamical two-Mode Squeezing of Thermal Fluctuations in a Cavity Optomechanical System," 
   \href{https://doi.org/10.1103/PhysRevLett.116.103601}{Phys. Rev. Lett. \textbf{116}, 103601 (2016)}.
		
\bibitem{ockeloen2016} C. F. Ockeloen-Korppi, E. Damsk{\"a}gg, J.-M. Pirkkalainen, T. T. Heikkil{\"a}, F. Massel, and M. A. Sillanp{\"a}{\"a}, ``Low-noise amplification and frequency conversion with a multiport microwave optomechanical device," 
   \href{https://doi.org/10.1103/PhysRevX.6.041024}{Phys. Rev. X \textbf{6}, 041024 (2016)}.
	
\bibitem{ockeloen20172} C. F. Ockeloen-Korppi, E. Damsk{\"a}gg, J.-M. Pirkkalainen, T. T. Heikkil{\"a}, F. Massel, and M. A. Sillanp{\"a}{\"a}, ``Noiseless quantum measurement and squeezing of microwave fields utilizing mechanical vibrations,"
   \href{https://doi.org/10.1103/PhysRevLett.118.103601}{Phys. Rev. Lett. \textbf{118}, 103601 (2017)}.
	
\bibitem{pirkkalainen2015} J.-M. Pirkkalainen, E. Damsk{\"a}gg, M. Brandt, F. Massel, and M. A. Sillanp{\"a}{\"a}, ``Squeezing of quantum noise of motion in a micromechanical resonator," 
   \href{https://doi.org/10.1103/PhysRevLett.115.243601}{Phys. Rev. Lett. \textbf{115}, 243601 (2015)}.
	
\bibitem{xong2018}H. Xiong and Y. Wu, ``Fundamentals and applications of optomechanically induced transparency,"
   \href{https://doi.org/10.1063/1.5027122}{Appl. Phys. Rev \textbf{5}, 031305 (2018)}.	
	
\bibitem{mikaeili2022} H. Mikaeili, A. Dalafi, M. Ghanaatshoar, and B. Askari, ``Ultraslow light realization using an interacting Bose–Einstein condensate trapped in a shallow optical lattice," 
   \href{https://doi.org/10.1038/s41598-022-08250-9}{Sci. Rep. \textbf{12}, 4428 (2022)}. 
	
\bibitem{agarwal2010} G. S. Agarwal and S. Huang, ``Electromagnetically induced transparency in mechanical effects of light," 
	\href{https://doi.org/10.1103/PhysRevA.81.041803}{Phys. Rev. A \textbf{81}, 041803 (2010)}. 
	
\bibitem{weis2010} S. Weis, R. Rivière, S. Deleglise, E. Gavartin, O. Arcizet, A. Schliesser, and T. J. Kippenberg, ``Optomechanically Induced Transparency," 
   \href{https://doi.org/10.1126/science.1195596}{Science \textbf{330}, 1520 (2010)}.
	
	
\bibitem{safavi2011} A. H. Safavi-Naeini, T. P. Mayer Alegre, J. Chan, M. Eichenfield, M. Winger, Q. Lin, J. T. Hill, D. E. Chang, and O. Painter, ``Electromagnetically induced transparency and slow light with optomechanics," 
   \href{https://doi.org/10.1038/nature09933}{Nature \textbf{472}, 69 (2011)}.
	
\bibitem{karuza2013} M. Karuza, C. Biancofiore, M. Bawaj, C. Molinelli, M. Galassi, R. Natali, P. Tombesi, G. Di Giuseppe, and D. Vitali, ``Optomechanically induced transparency in a membrane-in-the-middle setup at room temperature," 
   \href{https://doi.org/10.1103/PhysRevA.88.013804}{Phys. Rev. A \textbf{88}, 013804 (2013)}.
	
\bibitem{kronwald2013} A. Kronwald and F. Marquardt, ``Optomechanically induced transparency in the nonlinear Quantum Regime," 
   \href{https://doi.org/10.1103/PhysRevLett.111.133601}{Phys. Rev. Lett. \textbf{111}, 133601 (2013)}.
	
\bibitem{motazedifard2022} A. Motazedifard, A. Dalafi, and M. H. Naderi, ``Negative cavity photon spectral function in an optomechanical system with two parametrically-driven mechanical oscillators,"
   \href{https://doi.org/10.1364/OE.499409}{Opt. Exp. 31, 36615 (2023)}.
	
\bibitem{yang2019}X. Yang,ZH. Yin, and M. Xiao, ``Optomechanically induced entanglement,"
   \href{https://doi.org/10.1103/PhysRevA.99.013811}{Phys. Rev. A. \textbf{99}, 013811 (2019)}.			
			
\bibitem{palomaki2013} T. A. Palomaki, J. D. Teufel, R.W. Simmonds, and K.W. Lehnert, ``Entangling mechanical motion with microwave fields,"
    \href{https://doi.org/10.1126/science.1244563}{Science \textbf{342}, 710 (2013)}.
	
\bibitem{paternostro2007} M. Paternostro, D. Vitali, S. Gigan, M. S. Kim, C. Brukner, J. Eisert, and M. Aspelmeyer, ``Creating and probing multipartite macroscopic entanglement with light,"
   \href{https://doi.org/10.1103/PhysRevLett.99.250401}{Phys. Rev. Lett. \textbf{99}, 250401 (2007)}.
		
\bibitem{dalafi2018} A. Dalafi, M. H. Naderi, and A. Motazedifard, ``Effects of quadratic coupling and squeezed vacuum injection in an optomechanical cavity assisted with a Bose-Einstein condensate," 
   \href{https://doi.org/10.1103/PhysRevA.97.043619}{Phys. Rev. A \textbf{97}, 043619 (2018)}.
	
\bibitem{barzanjeh2019} S. Barzanjeh, E. S. Redchenko, M. Peruzzo, M. Wulf, D. P. Lewis, G. Arnold, and J. M. Fink, ``Stationary Entangled Radiation from Micromechanical Motion," 
   \href{https://doi.org/10.1038/s41586-019-1320-2}{Nature. \textbf{570}, 480 (2019)}.
	
	
\bibitem{mari2013} A. Mari, A. Farace, N. Didier, V. Giovannetti, and R. Fazio, ``Measures of quantum synchronization in continuous variable systems," 
   \href{https://doi.org/10.1103/PhysRevLett.111.103605}{Phys. Rev. Lett. \textbf{111}, 103605 (2013)}.
			
	
\bibitem{kippenberg2007} T. J. Kippenberg and K. J. Vahala, ``Cavity Opto-Mechanics," 
   \href{https://doi.org/10.1364/OE.15.017172}{Opt. Exp. \textbf{15}, 17172 (2007)}.
	
\bibitem{tsang2010} M. Tsang and C.M. Caves, ``Coherent quantum-noise cancellation for optomechanical Sensors," 
   \href{https://doi.org/10.1103/PhysRevLett.105.123601}{Phys. Rev. Lett. \textbf{105} 123601 (2010)}.
	
	
\bibitem{buchmann2016} L. F. Buchmann, S. Schreppler, J. Kohler, N. Spethmann, and D. M. Stamper-Kurn, ``Complex squeezing and force measurement beyond the standard quantum limit," 
   \href{https://doi.org/10.1103/PhysRevLett.117.030801}{Phys. Rev. Lett. \textbf{117} 030801 (2016) }.
	
\bibitem{moler2017} Ch. B. M\o ller, R. A. Thomas, G. Vasilakis, E. Zeuthen, Y. Tsaturyan, M. Balabas, K. Jensen, A. Schliesser, K. Hammerer, and E. S. Polzik, ``Quantum back-action-evading measurement of motion in a negative mass reference frame," 
  \href{https://doi.org/10.1038/nature22980}{Nature Phys. \textbf{547}, 191(2017)}.
	
\bibitem{liu2019} Sh. Liu, B. Liu, J. Wang, T. Sun, and W. X. Yang, ``Realization of a highly sensitive mass sensor in a quadratically coupled optomechanical system," 
   \href{https://doi.org/10.1103/PhysRevA.99.033822}{Phys. Rev. A \textbf{99}, 033822 (2019)}.
	
\bibitem{allahverdi2022} H. Allahverdi, A. Motazedifard, A. Dalafi, D. Vitali, and M. H. Naderi, ``Homodyne coherent quantum noise cancellation in a hybrid optomechanical force sensor,"
   \href{https://doi.org/10.1103/PhysRevA.106.023107}{Phys. Rev. A \textbf{106}, 023107 (2022)}.
	
\bibitem{brunelli2019} M. Brunelli, D. Malz, and A. Nunnenkamp, ``Conditional dynamics of optomechanical two-tone backaction-evading measurements," 
   \href{https://doi.org/10.1103/PhysRevLett.123.093602}{Phys. Rev. Lett. \textbf{123}, 093602 (2019)}.
	
	
	\bibitem{ockeloen2018} C. F. Ockeloen-Korppi, E. Damsk{\"a}gg, G. S. Paraoanu, F. Massel, and M. A. Sillanp{\"a}{\"a}, "Revealing Hidden Quantum Correlations in an Electromechanical Measurement," 
	\href{https://doi.org/10.1103/PhysRevLett.121.243601}{Phys. Rev. Lett. \textbf{121}, 243601 (2018)}.
		
\bibitem{motazedifardAVS2021} A. Motazedifard, A. Dalafi, and M. H. Naderi, ``Ultra-precision quantum sensing and measurement based on nonlinear hybrid optomechanical systems containing ultracold atoms or atomic-BEC," 
   \href{https://doi.org/10.1116/5.0035952}{ AVS Quantum Science \textbf{3}, 024701 (2021)}. 
	
\bibitem{ebrahimi2021} M. Ebrahimi, A. Motazedifard, and M. B. Harouni, ``Single-quadrature quantum magnetometry in cavity electromagnonics,"
   \href{https://doi.org/10.1103/PhysRevA.103.062605}{Phys. Rev. A \textbf{103}, 062605 (2021)}.
	
\bibitem{bemani2021} F. Bemani, O. Cernotík, L. Ruppert, D. Vitali, and R. Filip, ``Force sensing in an optomechanical system with feedback-controlled in-loop light,"
   \href{https://doi.org/10.1103/PhysRevApplied.17.034020}{Phys. Rev. Applied \textbf{17}, 034020 (2022)}.
   
 \bibitem{barzanjehQIlluminationOMS2015} Sh. Barzanjeh, S. Guha, Ch. Weedbrook, D. Vitali, J. H. Shapiro, and S. Pirandola, ``Microwave quantum illumination,"
  \href{https://doi.org/10.1103/PhysRevLett.114.080503}{Phys. Rev. Lett. \textbf{114}, 080503 (2015)}.
 
 \bibitem{barzanjeh2020} S. Barzanjeh, S. Pirandola, D. Vitali, and J. M. Fink, ``Microwave quantum illumination using a digital receiver"
   \href{https://doi.org/10.1126/sciadv.abb0451}{Sci. Adv. \textbf{6}, eabb0451 (2020)}.
   
\bibitem{courty2003} J. M. Courty, A.Heidmann, and M. Pinard, ``Quantum locking of mirrors in interferometers"
   \href{https://doi.org/10.1103/PhysRevLett.90.083601}{Phys. Rev. Lett. \textbf{90} 083601(2003)}.
  
\bibitem{kurizki2015}G. Kurizki, P. Bertet, Y. Kubo, K. Mølmer, D. Petrosyan, P. Rabl, and J. Schmiedmayer, ``Quantum technologies with hybrid systems,"
   \href{https://doi.org/10.1073/pnas.1419326112}{ Proceedings of the National Academy of Sciences \textbf{13}, 3866 (2015).}.
	
\bibitem{hafezi2022}M. Hafezi, Z. Kim, S. L. Rolston, L. A. Orozco, B. L. Lev, and J. M. Taylor, ``Atomic interface between microwave and optical photons,"
   \href{https://doi.org/10.1103/PhysRevA.85.020302}{Phys. Rev. A \textbf{85}, 020302(R) (2022)}.		
	
\bibitem{verdu2009}J. Verdú, H. Zoubi, Ch. Koller, J. Majer, H. Ritsch, and J. Schmiedmayer, ``Strong magnetic coupling of an ultracold gas to a superconducting waveguide cavity,"
   \href{https://doi.org/10.1103/PhysRevLett.103.043603}{Phys. Rev. Lett. \textbf{103}, 043603 (2009)}.	
	
\bibitem{imamoglu2009}A. Imamoğlu, ``Cavity QED based on collective magnetic dipole coupling: spin ensembles as hybrid two-level systems,"
   \href{https://doi.org/10.1103/PhysRevLett.102.083602}{Phys. Rev. Lett. \textbf{102}, 083602 (2009)}.	
	
\bibitem{marcos2010}D. Marcos, M. Wubs, J. M. Taylor, R. Aguado, M. D. Lukin, and A. S. Sørensen, ``Coupling nitrogen-vacancy centers in diamond to superconducting flux qubits,"
   \href{https://doi.org/10.1103/PhysRevLett.105.210501}{Phys. Rev. Lett. \textbf{105}, 210501 (2010)}.
   
\bibitem{safavi20112}A. H. Safavi-Naeini and O. Painter, ``Proposal for an optomechanical traveling wave phonon–photon translator,"
   \href{https://doi.org/10.1088/1367-2630/13/1/013017}{New J. Phys. \textbf{13}, 013017 (2011)}.
   
\bibitem{regal2011}C. A. Regal and K. W. Lehnert, ``From cavity electromechanics to cavity optomechanics,"
   \href{https://doi.org/10.1088/1742-6596/264/1/012025}{J. Phys.: Conf. Ser. \textbf{264}, 012025 (2011)}.
   
\bibitem{bochman2013}J. Bochman, A. Vainsencher, D. D. Awschalom, and A. N. Cleland, ``Nanomechanical coupling between microwave and optical photons.,"
   \href{https://doi.org/10.1038/nphys2748}{Nature Phys. \textbf{9}, 712 (2013)}.
  	
\bibitem{arnold2020}G. Arnold, M. Wulf, S. Barzanjeh, E. S. Redchenko, A. Rueda, W. J. Hease, F. Hassani, and J. M. Fink, ``Converting microwave and telecom photons with a silicon photonic nanomechanical interface,"
    \href{https://doi.org/10.1038/s41467-020-18269-z}{Nat. Commun \textbf{11}, 4460 (2020)}.
	
\bibitem{vitali2007}D. Vitali, P. Tombesi, M.J. Woolley, A.C. Doherty, and G.J. Milburn, ``Entangling a nanomechanical resonator and a superconducting microwave cavity,"
   \href{https://doi.org/10.1103/PhysRevA.76.042336}{Phys. Rev. A. \textbf{76}, 042336 (2007)}.

\bibitem{thompson2008}J. D. Thompson, B. M. Zwickl, A. M. Jayich, F. Marquardt, S. M. Girvin, and J. G. E. Harris, ``Strong dispersive coupling of a high-finesse cavity to a micromechanical membrane,"
   \href{https://doi.org/10.1038/nature06715}{Nature \textbf{452}, 72 (2008)}.

\bibitem{forsch2020}M. Forsch, R. Stockill, A. Wallucks, I. Marinković, C. Gärtner, R. A. Norte, F. van Otten, A. Fiore, K. Srinivasan, and S. Gröblacher, ``Microwave-to-optics conversion using a mechanical oscillator in its quantum ground state,"
   \href{https://doi.org/10.1038/s41567-019-0673-7}{Nat. Phys. \textbf{16}, 69 (2020)}.

\bibitem{barzanjeh2022}S. Barzanjeh, A. Xuereb, S. Gröblacher, M. Paternostro, C. A. Regal, and E. M. Weig , ``Optomechanics for quantum technologies,"
   \href{https://doi.org/10.1038/s41567-021-01402-0}{Nat. Phys. \textbf{18}, 15 (2022)}.

\bibitem{bonaldi2023}M. Bonaldi, A. Borrielli, G. D. Giuseppe, N. Malossi, B. Morana, R. Natali, P. Piergentili, P. M. Sarro, E. Serra, and D. Vitali, ``Low noise opto-electro-mechanical modulator for RF-to-optical transduction in quantum communications,"
    \href{ https://doi.org/10.3390/e25071087}{Entropy \textbf{25}, 1087 (2023)}.
    
\bibitem{eshaghi2022}N. Eshaqi-Sani, S. Zippilli, and D. Vitali, ``Nonreciprocal conversion between radio-frequency and optical photons with an optoelectromechanical system,"
    \href{https://doi.org/10.1103/PhysRevA.106.032606}{Phys. Rev. A \textbf{106}, 032606 (2022)}.
 
\bibitem{andrews2014}R. W. Andrews, R. W. Peterson, T. P. Purdy, K. Cicak,R. W. Simmonds , C. A. Regal and K. W. Lehnert, ``Bidirectional and efficient conversion between microwave and optical light,"
   \href{ https://doi.org/10.1038/nphys2911}{Nature Phys. \textbf{10}, 321 (2014)}.   	

\bibitem{haghighi2018}I. M. Haghighi, N. Malossi, R. Natali, G. Di Giuseppe, and D. Vitali, ``Sensitivity-bandwidth limit in a multimode optoelectromechanical transducer,"
   \href{https://doi.org/10.1103/PhysRevApplied.9.034031}{Phys. Rev. Appl. \textbf{9}, 03403 (2018)}.
   
\bibitem{karuza2012}M. Karuza, M. Galassi, C. Biancofiore, C. Molinelli, R. Natali, P. Tombesi, G. Di Giuseppe, and D. Vitali, ``Tunable linear and quadratic optomechanical coupling for a tilted membrane within an optical cavity: theory and experiment,"
   \href{https://doi.org/10.1088/2040-8978/15/2/025704}{J. Opt. \textbf{15}, 025704 (2012)}.  
   
\bibitem{khorasani2018}S. Khorasani,``Higher-order interactions in quantum optomechanics: analysis of quadratic terms"
   \href{https://doi.org/10.1038/s41598-018-35055-6}{Sci. Rep. \textbf{8}, 16676 (2018).}
   
\bibitem{javich2008}A. M. Jayich, J. C. Sankey, B. M. Zwickl, C. Yang, J. D. Thompson, S. M. Girvin, A. A. Clerk, F.Marquardt, and J. G. E. Harris, ``Dispersive optomechanics: a membrane inside a cavity"
   \href{https://doi.org/10.1088/1367-2630/10/9/095008}{New J. Phys. \textbf{10}, 095008 (2008)}.
   
\bibitem{purdy2010}T. P. Purdy, D. W. C. Brooks, T. Botter, N. Brahms, Z. Y. Ma, and D. M. Stamper-Kurn, ``Tunable cavity optomechanics with ultracold atoms,"
   \href{https://doi.org/10.1103/PhysRevLett.105.133602}{ Phys. Rev. Lett. \textbf{105}, 133602 (2010)}.
   
\bibitem{liao2014}J. Q. Liao and F. Nori, ``Single-photon quadratic optomechanics,"
   \href{https://doi.org/10.1038/srep06302}{Sci. Rep. \textbf{4}, 6302 (2014 )}.
   
\bibitem{kim2015}E. J. Kim, J. R. Johansson, and F. Nori, ``Circuit analog of quadratic optomechanics,"
\href{https://doi.org/10.1103/PhysRevA.91.033835}{Phys. Rev. A \textbf{91}, 033835 (2015)}.

\bibitem{vanner2011}M. R. Vanner,``Selective linear or quadratic optomechanical coupling via measurement,"
  \href{https://doi.org/10.1103/PhysRevX.1.021011}{ Phys. Rev. X \textbf{1}, 021011 (2011)}.
  
\bibitem{huang2011}S. Huang and G. S. Agarwal,``Electromagnetically induced transparency from two-phonon processes in quadratically coupled membranes,"
   \href{https://doi.org/10.1103/PhysRevA.83.023823}{ Phys. Rev. A \textbf{83}, 023823 (2011)}.
   
\bibitem{si2017}L. G. Si, H. Xiong, M. S. Zubairy, and Y. Wu,``Optomechanically induced opacity and amplification in a quadratically coupled optomechanical system,"
   \href{https://doi.org/10.1103/PhysRevA.95.033803}{Phys. Rev. A \textbf{95}, 033803 (2017)}.
   
\bibitem{liu2017}S. Liu, W. X. Yang, T. Shui, Z. Zhu, and A. X. Chen, ``Tunable two-phonon higher-order sideband amplification in a quadratically coupled optomechanical system,"
   \href{https://doi.org/10.1038/s41598-017-17974-y}{Sci. Rep. \textbf{7}, 17637 (2017)}.
   
\bibitem{liao2013}J.-Q. Liao and F. Nori, ``Photon blockade in quadratically coupled optomechanical systems,"
   \href{https://doi.org/10.1103/PhysRevA.88.023853}{Phys. Rev. A \textbf{88}, 023853 (2013)}.
 
\bibitem{shi2018} H. Q. Shi, X. T. Zhou, X. W. Xu, and N. H. Liu, ``Phonon blockade in a nanomechanical resonator quadratically coupled to a two-level system,"
   \href{https://doi.org/10.1038/s41598-019-45027-z}{Sci. Rep. \textbf{8}, 2212 (2018)}.

\bibitem{lu2018}X. Y. Lü, L. L. Zheng, G. L. Zhu, and Y. Wu,``Single-photon-triggered quantum phase transition,"
   \href{https://doi.org/10.1103/PhysRevApplied.9.064006}{ Phys. Rev. Appl. \textbf{9}, 064006 (2018)}.

\bibitem{zhan2013}X.-G. Zhan, L.-G. Si, A.-S. Zheng, and X. Yang, ``Tunable slow light in a quadratically coupled optomechanical system,"
   \href{https://doi.org/10.1088/0953-4075/46/2/025501}{J. Phys. B \textbf{46}, 025501 (2013)}.
   
\bibitem{clerk2010}A. A. Clerk, F. Marquardt, and J. G. E. Harris, ``Quantum measurement of phonon shot noise,"
   \href{https://doi.org/10.1103/PhysRevLett.104.213603}{Phys. Rev. Lett. \textbf{104}, 213603 (2010)}.  
   
\bibitem{buchmann2012}L. F. Buchmann, L. Zhang, A. Chiruvelli, and P. Meystre, ``Macroscopic tunneling of a membrane in an optomechanical double-well potential,"
   \href{https://doi.org/10.1103/PhysRevLett.108.210403}{Phys. Rev. Lett. \textbf{108}, 210403 (2012).}
   
\bibitem{nunnenkamp2010} A. Nunnenkamp, K. Børkje, J. G. E. Harris, and S. M. Girvin, ``Cooling and squeezing via quadratic optomechanical coupling," 
   \href{https://doi.org/10.1103/PhysRevA.82.021806}{Phys. Rev. A \textbf{82}, 021806(R) (2010)}
   
\bibitem{asjad2014}M. Asjad, G. S. Agarwal, M. S. Kim, P. Tombesi, G. Di Giuseppe, and D.Vitali, ``Robust stationary mechanical squeezing in a kicked quadratic optomechanical system,"
   \href{https://doi.org/10.1103/PhysRevA.89.023849}{Phys. Rev. A \textbf{89}, 023849 (2014).}
   
\bibitem{gu2015}W.-J. Gu, Z. Yi, L.-H. Sun, and D.-H. Xu, ``Mechanical cooling in single-photon optomechanics with quadratic nonlinearity,"
   \href{https://doi.org/10.1103/PhysRevA.92.023811}{Phys. Rev. A \textbf{92}, 023811 (2015).}
   
\bibitem{banerjee2023}P. Banerjee, S. Kalita, and A. K. Sarma, "Robust mechanical squeezing beyond 3 dB in a quadratically coupled optomechanical system"
   \href{https://doi.org/10.1364/JOSAB.483944}{J. Opt. Soc. Am. B \textbf{40}, 1398 (2023)}
   
\bibitem{jacobs2009}K. Jacobs, L. Tian, and J. Finn, ``Engineering superposition states and tailored probes for nanoresonators via open-loop control,"
   \href{https://doi.org/10.1103/PhysRevLett.102.057208}{Phys. Rev. Lett. \textbf{102}, 057208 (2009)}

\bibitem{shi2013}H. Shi and M. Bhattacharya, ``Quantum mechanical study of a generic quadratically coupled optomechanical system,"
   \href{https://doi.org/10.1103/PhysRevA.87.043829}{Phys.Rev. A \textbf{87}, 043829 (2013).}   
   
\bibitem{tan2013}H. Tan, F. Bariani, G. Li, and P. Meystre, ``Generation of macroscopic quantum superpositions of optomechanical oscillators by dissipation,"
   \href{https://doi.org/10.1103/PhysRevA.88.023817}{Phys. Rev. A \textbf{88}, 023817 (2013).}
   
\bibitem{abdi2016}M. Abdi, P. Degenfeld-Schonburg, M. Sameti, C. Navarrete-Benlloch, and M. J. Hartmann, ``Dissipative optomechanical preparation of macroscopic quantum superposition states,"
   \href{https://doi.org/10.1103/PhysRevLett.116.233604}{Phys. Rev. Lett. \textbf{116}, 233604 (2016).}
   
\bibitem{zhang2018}X. Y. Zhang, Y. H. Zhou, Y. Q. Guo, and X. X. Yi, ``Optomechanically induced transparency in optomechanics with both linear and quadratic coupling,"
   \href{https://doi.org/10.1103/PhysRevA.98.053802}{ Phys. Rev. A \textbf{98}, 053802 (2018).}
   
\bibitem{chao2021}S. L. Chao, Z. Yang, C. S.  Zhao, R. Peng, and L. Zhou, ``Force sensing in a dual-mode optomechanical system with linear–quadratic coupling and modulated photon hopping,"
   \href{https://doi.org/10.1364/OL.425484}{ Opt. Lett. \textbf{46}, 3075 (2021).}
   
\bibitem{xuereb2013}A. Xuereb, and M. Paternostro, ``Selectable linear or quadratic coupling in an optomechanical system,"
   \href{https://doi.org/10.1103/PhysRevA.87.023830}{ Phys. Rev. A \textbf{87}, 023830 (2013).}
   
 \bibitem{barzanjeh2011}Sh. Barzanjeh, D. Vitali, P. Tombesi, and G.J. Milburn, ``Entangling optical and microwave cavity modes by means of a nanomechanical resonator,"
   \href{https://doi.org/10.1103/PhysRevA.84.042342}{Phys. Rev. A. \textbf{84}, 042342 (2011)}. 
   
\bibitem{bhattacharya2008}M. Bhattacharya, H. Uys, and P. Meystre, ``Optomechanical trapping and cooling of partially reflective mirrors,"
   \href{https://doi.org/10.1103/PhysRevA.77.033819}{ Phys. Rev. A \textbf{77}, 033819 (2008)}. 
   
\bibitem{seok2013}H. Seok, L. F. Buchmann, E. M. Wright, and P. Meystre, ``Multimode strong-coupling quantum optomechanics,"
   \href{https://doi.org/10.1103/PhysRevA.88.063850}{Phys. Rev. A \textbf{88}, 063850( 2013)}.
   
\bibitem{gardiner2000} C. W. Gardiner and P. Zoller, \textit{Quantum Noise}
   \href{}{(Springer, Berlin, 2000)}.
   
 \bibitem{sainadh2015}S. Sainadh and M.A. Kumar, ``Effects of linear and quadratic dispersive couplings on optical squeezing in an optomechanical system,"
   \href{https://doi.org/10.1103/PhysRevA.92.033824}{Phys. Rev. A. \textbf{92}, 033824 (2015)}.  
 
\bibitem{teufel20112}J. D. Teufel, D.  Li, M. S. Allman, K. Cicak, A. J. Sirois, J. D. Whittaker, and R. W. Simmonds, ``Circuit cavity electromechanics in the strong-coupling regime,"
   \href{https://doi.org/10.1038/nature09898}{Nature \textbf{471}, 204 (2011)}. 
   
 \bibitem{dejesus1987}E. X. DeJesus and Ch. Kaufman, ``Routh-Hurwitz criterion in the examination of eigenvalues of a system of nonlinear ordinary differential equations,"
   \href{https://doi.org/10.1103/PhysRevA.35.5288}{Phys. Rev. A. \textbf{35}, 5288 (1987)}. 
   
\bibitem{lyapunov1992}A.M. Lyapunov, \textit{The General Problem of Stability of Motion}
   \href{https://doi.org/10.1080/00207179208934253}{(Taylor \& Francis, London, 1992).}

\bibitem{plenio2005}M. B. Plenio, ``Logarithmic negativity: A full entanglement monotone that is not convex,"
   \href{https://doi.org/10.1103/PhysRevLett.95.090503}{Phys. Rev. Lett. \textbf{95}, 090503 (2005)}.
   
\bibitem{vidal2002}G. Vidal, and R. F. Werner, ``Computable measure of entanglement,"
   \href{https://doi.org/10.1103/PhysRevA.65.032314}{Phys. Rev. A \textbf{65}, 032314 (2002)}. 
   
 \bibitem{adesso2004}G. Adesso, A. Serafini, and F. Illuminati, ``Extremal entanglement and mixedness in continuous variable systems,"
   \href{https://doi.org/10.1103/PhysRevA.70.022318}{Phys. Rev. A \textbf{70}, 022318 (2004)}.   
   
\bibitem{simon2000}R. Simon,``Peres-Horodecki separability criterion for continuous variable systems, "
   \href{https://doi.org/10.1103/PhysRevLett.84.2726}{ Phys. Rev. Lett. \textbf{84}, 2726 (2000)}.   
  
\bibitem{adesso2005}G. Adesso and F. Illuminati, ``Equivalence between entanglement and the optimal fidelity of continuous variable teleportation,''
   \href{https://doi.org/10.1103/PhysRevLett.95.150503}{Phys. Rev. Lett. \textbf{95}, 150503 (2005)}.
   
\bibitem{mari2008}A. Mari and D. Vitali, ``Optimal fidelity of teleportation of coherent states and entanglement,'' 
   \href{https://doi.org/10.1103/PhysRevA.78.062340}{Phys. Rev. A \textbf{78}, 062340 (2008)}.
   
\bibitem{fan2023}Z.-Y. Fan, L. Qiu, S. Gröblacher, and J. Li, ``Microwave-optics entanglement via cavity optomagnomechanics,'' 
   \href{https://doi.org/10.1002/lpor.202200866}{Laser Photon. Rev \textbf{17}, 2200866 (2023)}.
   
\bibitem{rueda2019}A. Rueda, W. Hease, S. Barzanjeh, and J. M. Fink, ``Electro-optic entanglement source for microwave to telecom quantum state transfer,'' 
   \href{https://doi.org/10.1038/s41534-019-0220-5}{npj Quantum Inf \textbf{5}, 108 (2019)}. 
   
\bibitem{zheng2024}Q. Zheng, W. Zhong, G. Cheng, and A. Chen, ``Nonreciprocal microwave-optical entanglement in a magnon-based hybrid system,'' 
   \href{https://doi.org/10.1063/5.0190162}{J. Appl. Phys. \textbf{135}, 084401 (2024)}. 
   
\bibitem{sahu2023}R. Sahu, L. Qiu, W. Hease, G. Arnold, Y. Minoguchi, P. Rabl, and J. M. Fink, ``Entangling microwaves with optical light,'' 
   \href{https://doi.org/10.1126/science.adg3812}{Science \textbf{380}, 718 (2023)}. 
   
\bibitem{barzanjeh20112}Sh. Barzanjeh, M. H. Naderi, and M. Soltanolkotabi, ``Back-action ground-state cooling of a micromechanical membrane via intensity-dependent interaction,"
   \href{https://doi.org/10.1103/PhysRevA.84.023803}{Phys. Rev. A \textbf{84}, 023803 (2011)}.
   
   \bibitem{shahidani2014}S. Shahidani, M. H. Naderi, M. Soltanolkotabi, and S. Barzanjeh, ``Steady-state entanglement, cooling, and tristability in a nonlinear optomechanical cavity,"
   \href{https://doi.org/10.1364/JOSAB.31.001087}{J. Opt. Soc. Am. B \textbf{31}, 1087 (2014)}.
   
   
   \bibitem{wang2023}L. Wang, W. Zhang, Sh. Liu, Sh. Zhang, and H.F. Wang, ``Simultaneous cooling and synchronization of the mechanical and the radio-frequency resonators via voltage modulation,"
   \href{https://doi.org/10.1140/epjqt/s40507-023-00191-0}{EPJ Quantum Technol. \textbf{10}, 34 (2023)}.
  
   \bibitem{genes20081}C. Genes, D. Vitali, P. Tombesi, S. Gigan, and M. Aspelmeyer, ``Ground-state cooling of a micromechanical oscillator: Comparing cold damping and cavity-assisted cooling schemes,"
     \href{https://doi.org/10.1103/PhysRevA.77.033804}{Phys. Rev. A \textbf{77}, 033804 (2008)}.
   
   
\bibitem{agarwal2016}G. S. Agarwal, and S. Huang, ``Strong mechanical squeezing and its detection,"
   \href{https://doi.org/10.1103/PhysRevA.93.043844}{Phys. Rev. A \textbf{93}, 043844 (2016)}.
   
\bibitem{lei2016}C. U. Lei, A. J. Weinstein, J. Suh, E. E. Wollman, A. Kronwald, F. Marquardt, A. A. Clerk, and K. C. Schwab, ``Quantum nondemolition measurement of a quantum squeezed state beyond the 3 dB limit,"
   \href{https://doi.org/10.1103/PhysRevLett.117.100801}{Phys. Rev. Lett. \textbf{117}, 100801 (2016)}.
   
\end{thebibliography}
\end{document}